\documentclass[onecolumn]{aa} 
\usepackage{graphicx}
\usepackage{txfonts}
\begin{document}
\title{Active Galactic Nuclei Jets and Multiple Oblique Shock Acceleration: Starved Spectra}


\author{A. Meli
          \inst{1}
          \and
          P. L. Biermann\inst{2,3,4,5,6}
         }

\institute{IFPA, Department of Astrophysics and Geophysics, University of Liege, Belgium\\
              \email{ameli@ulg.ac.be}
   \and
MPI f\"ur Radioastronomie, Bonn, Germany\\
$^{3}$ Inst. Exp. Nucl. Phys., Karlsruher Institut f{\"u}r Technologie KIT,
Germany   \\
$^{4}$ Department of Physics \& Astronomy, University of Alabama, 
Tuscaloosa, AL, USA\\
$^{5}$ Department of Physics, University of Alabama at Huntsville, AL, USA\\
$^{6}$ Department of Physics \& Astronomy, University of Bonn, Germany \\ 
             \email{plbiermann@mpifr-bonn.mpg.de}
             }



\abstract
{Shocks in jets and hot spots of Active Galactic Nuclei (AGN) are one prominent class of possible sources of very high energy cosmic
ray particles (above $10^{18}$eV). 
Extrapolating their spectrum to their plausible injection energy from some shock, implies an 
enormous hidden energy for a spectrum of index $\sim -2$. Some analyzes 
suggest the particles' injection spectrum at source to be as steep as -2.4 to -2.7, 
making the problem much worse, by a factor of order $10^{6}$.
Nevertheless, it seems implausible that more than at the very best 1/3 of the jet energy,
goes into the required flux of energetic particles thus, one would need to allow 
for the possibility that there is an energy problem, which we would like to 
address in this work.}
{Sequences of consecutive oblique shock features, or conical shocks, have been theorized and eventually observed in many AGN jets.
Based on that, we use by analogy the 'Comptonisation' effect and we propose a scenario of a single injection of particles  which 
are accelerated consecutively by several oblique shocks along the axis of an AGN jet.}
{We use detailed test-particle approximation Monte Carlo simulations in order to calculate particle spectra by acceleration at 
such a shock pattern while monitoring the efficiency of acceleration, calculating differential spectra.}
{We find that the first shock of a sequence of oblique shocks, establishes a low energy power-law spectrum with $\sim E^{-2.7}$.
The following consecutive shocks push the
spectrum up in energy, rendering flatter distributions with steep cut-offs, and characteristic depletion at low energies,  
an effect which could explain the puzzling apparent extra source power.}
{Our numerical calculations show a variation of spectral indexes, starved spectra and a general spectral flattening,  connecting 
to multiple shock acceleration conditions, the relativistic 
nature of the shocks and the steepness of the magnetic field to the shock normal, shedding further light into understanding the jet-magnetic field 
geometry and the irregular or flat spectra observed in many AGN jets (e.g. Cen A, 3C279, PKS 1510-089).
Furthermore, the $E^{-2.4}-E^{-2.7}$ UHECRs injected source spectra claimed by many authors,
could be explained by the superposition of several, perhaps many 
emission sources, all of which end their particle shock acceleration sequence with flatter, starved spectra produced by two or more consecutive 
oblique shocks along their jets, or could also imply a mixed component of the accelerated particles above $10^{19}$eV.}


\keywords{acceleration of particles, simulations}

\maketitle
%

\section{Introduction}

The flux of ultra high energy cosmic rays (UHECRs) at $E\ge10^{9}$~GeV 
is believed to arise in plasma shock environments in extragalactic sources; most favourable
being the hot spots or strong shocks in the jets and 
radio lobes of Active Galactic Nuclei (AGN) (e.g. Ginzburg \& Syrovatskii 1963, 
Biermann \& Strittmatter 1987, Rachen \& Biermann 1993) or alternatively, shocks in Gamma Ray Burst
environments (e.g. qualitatively by Biermann 1994, quantitatively by  Waxman 1995, Vietri 1995). Nevertheless,
An absence of neutrinos associated with cosmic-ray acceleration Gamma Ray Bursts (IceCube Collaboration, 2012)
shows that no neutrinos are associated with GRBs

In the mechanism of diffusive shock acceleration, particles repeatedly
gain energy in multiple crossings of an astrophysical shock discontinuity, due
to collisions with upstream and downstream magnetic scattering centers, 
resulting in a power-law spectrum extending up to the energy of the observed UHECRs events.
Initially, theoretical developments on particle acceleration were made by Fermi (1949, 1954),
Darwin (1954) and during 1970s, the work of Axford et al. (1977), Krymsky (1977),
Bell (1978) and Blandford \& Ostriker (1978) established the 1st order Fermi mechanism for 
diffusive particle shock acceleration.


Ultra-high energy measurements by the Auger experiment (Auger Collaboration 2007, 2010a, 2010b), 
have shown  that AGN seem a quite plausible candidate source for UHECRs.
Auger also reported a suppression of cosmic ray flux above $4 \cdot 10^{19}$ eV. The HiRes experiment reported 
an observation of the cutoff (HiRes Collaboration 2009). Moreover Auger measurements indicated 
a weak statistical trend which supports the idea that very high energy events are localized (Auger Collaboration 2010b).
In addition, the detection of very hard TeV-energy spectra of 
high-redshift ($z>0.15$) Blazars, like the case of 3C279 (Teshima et al., 2009) detected by the MAGIC telescope
at a redshift of $z=0.536$, poses  one of the most interesting questions in astroparticle physics and shock acceleration efficiency.

It seems implausible that more than at the very best of 1/3 of an AGN jet energy goes 
into energetic particles (Falcke and Biermann 1995, Falcke et al. 1995); and so equating the power in UHECRs, adding 
the additional power hidden at lower energies, and equating it with 
1/3 of the observed jet-power, it yields a condition on the spectrum:  that in turn this would imply a spectrum flatter than $E^{-2.2}$.
Thus, one needs to allow for the possibility that there is an energy problem, which we would 
like, among other aspects, to address in the present study.

The best simple power-law spectral fit to the UHECR ($\leq 10^{9}$ GeV) data suggests a spectral index between -2.4 and -2.7
(de Marco \& Stanev, 2005, Berezinsky et al. 2006, 2009). Thus a total power of $10^{48.5} \times D_{10}^{2}$ erg/s is required, 
where $D_{10}$ denotes the distance in units of 10 Mpc. 
The prime candidates seem to be Cen A and M87 
(Ginzburg \& Syrovatskii 1963, Biermann \& Strittmatter 1987). 
Very recently, Gopal-Krishna et al. (2010) obtained an excellent fit for an UHECR signal contribution by Cen A,
but not just a simple power-law, but a spectrum with a kink downwards.
Assuming that up to 1/3 of the total jet 
power can be supplied, then the maximal power possible to be 
provided  in UHECR is $10^{42.5}$ erg/s for Cen A, and $10^{44.5}$ erg/s for M87 
(e.g. Whysong \& Antonucci 2003, Abdo et al. 2009, Gopal-Krishna et al. 2010), which
falls far below than what is required to explain the 
data.  Therefore, one solution would be to consider spectra, that are maybe 
starved at lower energies.
We note here that, it seems plausible to assume that at around 
a few $10^{19}$ eV there are many candidates and only at higher energies
there are either very few or perhaps only one source contribution, which would worsen the energy problem 
(i.e. the distance of either Cen A or M87 and the interaction with the microwave background is not quite so important, 
see Greisen 1966, Zatsepin and Kuzmin 1966).


The observed standing shock features in AGN jets (e.g. 
Sanders 1983, Gawthorne 2006, Marscher et al. 2008), and the system of repeated  shocks observed in jet 
structure at vastly discrepant spatial resolutions,
like in the radio galaxy NGC6251 (Bridle and Perley, 1984) motivated
us for this work.  
Moreover, it is well known from normal quantum statistics, that prescribing the
number of photons in a box filled with hot gas of temperature $T$, their
total energy, and forbidding both creation and destruction of photons,
gives highly distorted Planck spectra. These spectra as a function of
photon energy $\epsilon$ can be described using a "chemical potential",
$\mu$ (e.g. Leighton 1959),  $N(\epsilon) \; \sim \; e^{-\mu -\epsilon/kT}$.
This is called incomplete Comptonisation effect.
The process of incomplete Comptonisation in the disks of accreting compact objects is one 
well known astrophysical application (e.g. Sunyaev 1970, Katz 1976, Rybicki and Lightman 1993), see figure 1.
If there is an insufficient number of photons available, the generated spectrum
is depressed at low energies. This entails, that a low energy extension
of high energy spectra may not be realistic, and therefore, we need to allow for
depressed low energy spectra, thus we wish to
draw here an analogy to the possibility that cosmic ray 
spectra produced by acceleration in multiple oblique shock structures
in AGN jets, may also be depressed (starved) in lower energies while extending to higher energies.
In the present study, we will show that the first shock from a shock sequence, 
establishes a power-law spectrum with a spectral index of $\sim 2.7$, while the following shocks of the sequence
push the particle spectrum up in energy, with flatter distributions manifesting a characteristic flux depletion at lower energies.

\subsection{Jets and multiple-oblique shocks}


First attempts to simulate relativistic hydrodynamical jets were made by Marti et al. (1994) and by Duncan and Hughes (1994)
for low Mach numbers and by Marti et al. (1995) for high Mach numbers. Their studies showed the
same trends with the global phenomenology of non-relativistic jet evolution, nevertheless it was
shown that relativistic jets propagate more efficiently into the ambient external medium comparing to a non-relativistic head jet velocity.
A year later, Massanglia et al. (1996a) presented more detailed two-dimensional simulations. 
Three-dimensional simulations (e.g. Arnold and Arnett 1986, Clarke 1996) also reveal internal 
oblique shock structures in light or moderately supersonic jets. Among many
insightful findings regarding the structure of a relativistic jet, it was found that
when a high pressure cocoon develops, it squeezes the jet and drives shock waves into it, which reflect 
on the axis and form a conical (in other words an oblique shock sequence) shape. The aspect of this interaction depends on the Mach number.
Furthermore, one could establish a relationship between an optimum angle ($a$) of the cone between its tangential 
shock surface and the flow parallel to the jet axis expressed in Lorentz factor $\Gamma$ as, $sin~ $a$~\propto 1/\Gamma$.
The critical parameter was found to be the inclination of the conical shocks that determined
the thrust behind the jet head, meaning in planar jet geometry: \textit{Oblique shocks must have a small inclination angle to the axis
in order to produce a strong acceleration effect}.

The physics of magneto-hydrodynamic (MHD) shocks topology in magnetized flows is still not well understood 
in the plasma and astrophysical community.  Efforts to understanding the jet physics via MHD simulations in 
1D, 2D and 3D are currently under investigation.
Despite the fact that jet dynamics is inherently three-dimensional (3D), much of the physics of these jets
can be obtained from simple 1D simulations of flow along a cylindrical shell.
For example Cheung et al. (2007) demonstrated that observed proper
motions of forward (superluminal) and reverse (subluminal)
knots can be reproduced precisely by a 1D relativistic MHD simulation model.
A very interesting work was presented in a series of papers by
Jones et al. (2001), Tregillis et al. (2001ab, 2004), O'Neil et al. (2006), etc where
the authors have conducted extensive studies of high-resolution 2D and 3D MHD simulations of supersonic jets, 
exploring the influence of the jet Mach number and the ambient medium on jet propagation and energy 
deposition over long distances.  Obviously, only 3D simulations can hope to
reproduce the obliquity of the features in some jets observed with very
high angular resolution, e.g. NGC6251, 3C66B, M87, etc

During the course of this work, we will assume a sequence of oblique shocks along the jet 
axis (as depicted in figure 4), following the \textit{reconfinement} mechanism (Sanders, 1983).
Sanders showed by the method of characteristics, that the reconfinement process is a 
typical property of jets observed transversely. The internal pressure in a free jet decreases 
rapidly with distance from the AGN nucleus, and therefore the jet would be expected to come into 
pressure equilibrium with the ambient medium. 
The reconfinement shocks allow the jet flow to adjust to the outside pressure staying supersonic.
These shocks shocks are almost periodically repeated, contacting the jet circumference and
they are mostly subluminal as long as the flow stays supersonic.

The  reconfinement mechanism and the solutions of a hydrodynamic wind (e.g. Fukue \& Okada 1990) show 
that a jet flow goes through multiple critical points, passing from subsonic to
supersonic and vice versa several times, with formation of shock structures. This mechanism is actually 
an attempt to recover the strength of a weak jet flow, which would mean that one would have repeatedly 
formations of superluminal to subluminal shocks and so forth, until the critical point were 
the outside pressure totally overpowers the pressure of the jet in the deceleration phase.




Observations of polarized synchrotron radiation in AGN jets (e.g. Agudo et al., 2001),
indicate stationary conical components in the observer frame in the parsec-scale 
jets, which could be identified as recollimation shocks. We note that systems of repeated oblique (conical) shocks are observed 
in jet structures at vastly discrepant spatial resolutions, like in the radio galaxy NGC6251 (Bridle \& Perley 1984), 
interpreting this as a quasi-self-similar pattern, with the local scales all proportional to radial distance
along the jet.
Marscher et al. (2008, 2010) showed high-resolution radio images and optical polarization measurements 
of the BL Lac type object PKS 1510-089, revealed a bright feature in its jet that caused a double flare of radiation from optical 
frequencies to TeV gamma-ray energies, as well as a delayed outburst at radio wavelengths. It was concluded that
the observational event started in a region with a \textit{helical} magnetic field which was identified with the acceleration and 
collimation zone predicted by theoretical works, and the variability in the observed brightness was due to 
emission by the plasma excited by a conical standing shock wave.
Furthermore, observations of the jet structure of M87 (Walker et al., 2008) and PKS 1510-089 (Marscher et al., 2008) 
near the central black hole indicate that, while there
can be moving shocks between 10 and 1000 Schwarzschild-radii ($r_s$), the first
strong stationary shock occurs at $\sim$ 3000 $r_s$, 
as already proposed by Markoff et al. (2005) and confirmed by
Britzen et al. (2008) in the case of the BL Lac type object S5 1803+784.


The evidence of helical jets transverse rotation
measure gradients across AGN jets, was predicted by Blandford (1993), and  
among others found by Gabuzda, Murray and Cronin (2004) in BL Lacs. 
Brown et al. (2009) found similar evidence in an the FRI radio galaxy 3C78.
Asada et al. (2008)  performed multi-frequency polarimetry for the quasar 
NRAO 140 using the VLBA. The observations revealed the existence of helical magnetic components 
associated with the jet itself.
Evidence for a co-existence of shocks and helical magnetic fields is given by
Keppens et al. (2008). 
It is important to note that using MHD simulations they explored the morphology of AGN jets by 
studying propagation characteristics of
a series of highly relativistic, helically magnetized jets and among other they showed that the magnetic 
helicity changes at internal cross-shocks, which then act to repeatedly re-accelerate the jet. 
Moreover, Nakamura et al. (2010), showed that extragalactic jets are the result of MHD shocks produced in helically
twisted, magnetized relativistic outflows.

Additionally, we note that state-of-the art laboratory experiments of supersonic highly conductive plasma flows
(e.g. Lebedev 2005, Ampleford et al. 2008) have also shown development of conical shocks forming at its base and along 
the axis of the developed jet.

The topology we assume here considers the axisymmetric reconfinement of shocks 
as a sequence of shocks with inclination angles $a, b, c, d$ within a supersonic AGN jet, with a helical magnetic field,
as shown in figure 4.
Specifically we assume that particle acceleration can take place in such internal shocks
as theorized or observed in e.g. PKS 1510-089, NGC6251, etc.

The paper is structured as follows: In section 2 we present our numerical method,
and we discuss results of individual and multiple shock acceleration studies. 
In section 3 we conclude discussing our findings.

\section{Simulations and results}

\subsection{Numerical method}

We have extended the relativistic particle shock acceleration code by Meli and Quenby (2003b)
to \textit{multiple} oblique shocks,
see topology shown in figure 5.  In principle, a Monte Carlo code gives a solution 
to the time independent Boltzmann equation by the following

\begin{equation}
\Gamma(V+\upsilon\mu )\frac{\partial f}{\partial x}=\frac{\partial f}{\partial t}\arrowvert_{c}\,,
\label{boltzmann:equ}
\end{equation}

where a steady state is assumed in the shock rest frame, $V$ is the fluid velocity, $\upsilon$ the velocity of the
particle, $\Gamma$ is the Lorentz factor of the fluid frame, $\mu=\cos\theta$ the cosine of the particle's pitch angle $\theta$
and $\left.\partial f/\partial t\right|_{c}$ is the collision operator.

The use of a Monte Carlo technique to solve equation 1, is dependent on the 
assumption that the collisions represent scattering in pitch angle and that the scattering 
is elastic in the fluid frame where there is no residual electric field. 
Since it is assumed that the Alfv{\'e}n waves have lower speed than the plasma
flow itself,  the scattering is elastic in the fluid frame. A phase averaged distribution 
function is appropriate to the diffusion approximation we employ, which uses many small angle scatterings.
The first order Fermi (diffusive) acceleration is  simulated provided there is a shock front, where the particles' 
guiding-center undergoes consecutive scatterings with the assumed magnetized media. In each shock crossing 
particles gain an amount of energy prescribed by the appropriate jump-condition equations.

A justification for the test-particle approach, is
the work of Bell (1978a,b) which has shown that
'thin' sub-shocks appear even in the non-linear regime, so at some energy above the plasma $\Gamma$
value, the accelerated particles may be dynamically unimportant while they re-cross the discontinuity.
Another way of arriving at the test-particle regime is to inject particles
well above the plasma particle energy when they are dynamically unimportant and
thus require the seed particles to have already been pre-accelerated. Particles are injected upstream towards the
shock and they are allowed to scatter in the respective fluid rest frames with their basic motion 
described by  a guiding center approximation.

In a shocked environment, flow into and out of the shock discontinuity is not along the shock 
normal (Begelman \& Kirk 1990),  
but a transformation is possible into the so called normal shock frame (NSH) to render 
the flow along the normal. 
Furthermore, an important Lorentz transformation from the NSH frame, to the so called
de Hoffmann-Teller frame (HT) (de Hoffmann \& Teller 1950) can apply.
In the HT frame the electric field {\bf E} = 0. Thus, one can study the diffusive shock acceleration mechanism in an 'electric-field-free' 
reference frame, boosting from the NSH frame by a speed $\beta_{HT}$ along the shock surface as, $\beta_{HT}\leq \beta_{NSH} \cdot \tan\psi$.
By inspecting this equation, given relativistic shocks, it becomes obvious that $\beta_{NSH} \sim 1$ for
all angles smaller than $\tan\psi = 1$, otherwise velocity $\beta_{HT}$ will be greater than one.
This physical causality gives rise to a classification of relativistic shocks into two categories: 
i) the so called \textit{subluminal} shock, when its inclination is $\tan\psi \leq 1$ 
(for these 'low-inclination' relativistic shocks the first order Fermi (diffusive) 
acceleration applies in the 'electric-field-free' HT frame). ii)
The \textit{superluminal} shock  when its inclination is $\tan\psi > 1$ (in superluminal shocks particles 
are accelerated by the so called, shock-drift acceleration mechanism in the presence of the electric field, see Armstrong \& Decker, 1979).

Standard theory poses the conservation of the first
adiabatic invariant in the HT frame in order to determine reflection or transmission of the particles. 
Reflection of particles during diffusive acceleration is important since it contributes
to the overall efficiency of acceleration by frame transformations.
In the HT frame the allowed and forbidden angles for transmission depend only on the input pitch and phase, 
not on rigidity, thus the results of Hudson (1965) apply in our model. 
In the relativistic shock situation anisotropy renders the input to 
the shock from upstream very anisotropic in pitch angle, but as was shown in Hudson (1965) and Meli (2003), 
it is an acceptable approximation to randomize phase before transforming to the HT frame and then to use the adiabatic 
invariant to decide on reflection/transmission.
While in the subluminal case, particle transmission at the shock can be decided in the HT frame employing
conservation of the first adiabatic invariant, in the superluminal case computations are followed
entirely in the fluid rest frames with reference to the NSH frame simply employed 
to check whether upstream or downstream shock conditions apply (Meli \& Quenby 2003b).
On the other hand, for the superluminal shock conditions, where the physical picture of the shock-drift acceleration mechanism applies,
it is necessary to abandon the guiding center approximation when the trajectories begin to intersect 
the shock surface. For this case, we consider a helical trajectory motion of each test-particle of momentum $p$,  
in the fluid frame, upstream or downstream, where the velocity coordinates ($u_x, u_y, u_z$) 
of the particle are calculated in 3-dimensional space (e.g. Meli \& Quenby 2003b).
Begelman and Kirk (1990)  pointed out, that in the blast 
wave frame the turbulence can be isotropic
and many shock stationary frame configurations can be superluminal.
Nevertheless, many polarization observations show that chaotic magnetic fields prevail at 
distances larger than a few parsecs, but they should be statistically anisotropic 
to produce a net linear polarisation as discussed in Korchakov \& Syrovatskii (1962).
To this end, Laing (1980) has pointed out that a chaotic magnetic field being initially 
isotropic becomes anisotropic after crossing a shock front occurred due to compression of plasma. 
Moreover, Meli \& Quenby (2003b) 
showed that a transformation from an initially
isotropic rest frame distribution to an accelerated flow frame leads to a co-moving
relativistic plasma frame field distribution lying close to the flow vector.
This condition allows for a range of subluminal situations when viewed in the shock frame.


The basic coordinate system to describe a shock is a Cartesian system $xyz$, where the shock plane lies on the
$yz$ plane. In principle a shock is placed at $x=0$, while $x<0$ corresponds to the upstream region and $x>0$ to the downstream one.
The direction of the flow in the shock rest frame is in the positive direction that is, from upstream to downstream. 
The reference frames used during the simulations
are the upstream and downstream fluid rest frames, the NSH frame and the HT frame.
Viewing in the NSH frame (see figure 3) and assuming that the flow velocity vector is parallel to the shock 
normal on sees that $i + a = 90^{o} $ and $\psi + i = 90^{o}$ thus $\psi = a$, which means that the inclination of the 
shock surface to the flow consequently indicates the inclination of the shock to the jet axis. 
Therefore, as long as $\psi$ is less or equal to $45^{o}$, the HT frame transformation is possible (figures 2, 3).
The scattering operator in our simulations is treated via a pitch angle scattering approach, see  Appendix and Meli et al. (2008). 

We allow  particles with pitch angle $\delta \theta$ chosen at random within  
$1/\Gamma \leq \delta \theta \leq 10/\Gamma$  (where $\Gamma$ is the shock's Lorentz factor). 
Furthermore, we allow $\lambda=10 r_g$ ($r_g$ the particle's gyroradius), which implies a mild turbulence, which is justifiable since 
there are mostly low magnetized tenuous plasmas in AGN jets further than 3000 Schwarzschild-radii. On turbulence media and 
particle scattering kinematics used in this work, see Appendix.

In the present model we assume four consecutive oblique shock planes e.g., 
with half-opening angles $a, b, c, d$ in respect to the axis, $x$. 
The magnetic field strength is set to B =$10^{-4}$ Gauss, typical for AGN jets. 
In figure 5 one sees the overall jet topology and shock geometries 
(not to scale) as the framework of our simulation. In the present model we assume that the overall area of the four 
shocks occupies a space of 100 pc, given most AGN jets have maximum lengths of a few hundreds pc.
We mention here that a particle's gyroradius,  of e.g. an energy $\sim 10^{11}$ GeV, 
within a typical jet magnetic field strength of $B \sim 10^{-4}$ Gauss, will be $\leq$ 100 pc.   
The particles differential spectrum is recorded downstream at a distance, $d=25$pc, from each shock, as shown
in figure 5. The energy spectra are calculated in the shock frame downstream at $d=25$pc from 
each shock and normalized to GeV units, assuming protons 
as the primary accelerated population. Moreover, particles leave the system if they reach a 
maximum momentum boundary set at $E_{max}=10^{11.5}$ GeV.


We start the simulation off the NSH frame as previously described and
a Lorentz transformation allocates the magnetic field vector perpendicular to the jet axis, therefore with 
an inclination to the shock normal, see figure 3.
We assume that the first shock of the sequence occurs at about 3000 Schwarzschild-radii from the AGN black hole 
(Markoff et al. 2005, Becker \& Biermann 2009).
Injection of a fixed number $N_i$ of particles takes place upstream towards the first shock.
The key point here is that in the following three shocks of the sequence, acceleration occurs for the same number of 
particles without an injection of new particles, as an analogy to the incomplete Comptonisation effect (Leighton, 1959), aforementioned
in section 1.

Particles are assigned a random phase so that a 3-dimensional transformation of 
momentum vectors can be achieved between the fluid and HT frames.
Away from the shock, the guiding center approximation is used so that
a test-particle moving a distance, $d$, along a field line at $\psi$ to the shock normal,
in the plasma frame has a probability of collision within $d$ given by $P(d)=1-\exp(-d/\lambda)=R$,
where the random number $R$ is $0 \leq R \leq 1$. Weighting the probability by the current in the 
field direction $\mu = cos\theta$ yields $h=-\lambda \mu \ln R$. 
The pitch angle is measured in the local fluid frame, while a value $x_i$ gives the distance 
of the particles to the shock front,  where the shock is assumed to be placed at $x=0$ as we mentioned.
Furthermore, $x_{i}$ is defined in the shock 
rest frame and the model assumes variability in only one spatial dimension.

Particle splitting was used to improve statistics, meaning when a particle reached a certain energy, it is replaced by 
two daughter-particles, which are otherwise identical to the primary particle, but having only half of the 
original statistical weight.   
We assume that there is no significant escape of particles perpendicularly the jet axis, off the
reconfinement boundary, see figure 5, as the particles remain inside the supersonic channel propagating to the right
along the x-axis.
This is justifiable by the mechanism of shear jet boundary acceleration (e.g. Earl et al. 1988, 
Stawarz and Ostrowski 2002, Rieger and Duffy 2006, Stawarz and Petrosian 2008, Sahayanathan 2009) which ensures 
that there is not a significant escape of particles off the jet reconfinement boundaries, which further implies that 
this mechanism enhances the attainment of very high energies.

One would consider that every particle accelerating from one shock to another in the four-shock sequence,
will reach the next shock where it can be further accelerated. Nevertheless, in a more realistic situation some particles
found downstream of the 4-th shock, may have been accelerated
in less than 4 shocks, therefore we allow here a fixed escape probability $P_{e}$ 
to be applied between the shocks during the acceleration process.
This escape probability gives actually the fraction of particles of the downstream
distribution of each shock that will not be further accelerated.
These particles remain in the system and contribute directly to the downstream distribution
of the 4-th shock.
In the simulations shown we choose  $P_{e}=0.3$ in order to simulate more realistically the
decompression and shock extension between shocks in a repetitive sequence  (e.g. Melrose and Pope, 1993). 
We note that a $P_{e}=1$ means that particles are produced by a single shock, while the case of $P_{e}=0$ shows that
all particles are transported through the subsequent shocks.

For an extensive description of the numerical approach and particle kinematics the reader is 
referred to Meli \& Quenby (2003a,b), Meli et al. (2008) and Appendix.

\begin{figure}[t]
\begin{center}
\includegraphics [height=7cm, width=6cm]{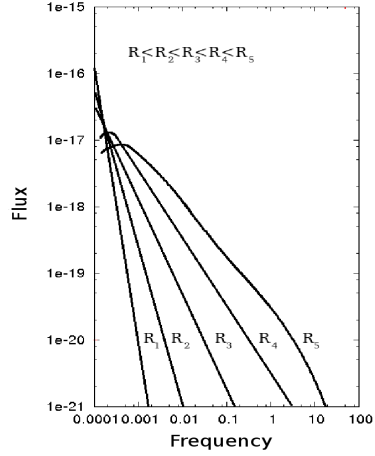}
\caption{\small{Radiation from a non-relativistic Compton 
photon scattering in a finite
gas sphere, where $R_i$ are for different gas radii. Figure reproduced based on Katz (1976).}}
\end{center}
\end{figure}

\begin{figure}[t]
\begin{center}
\includegraphics [height=8cm, angle=270, width=8cm]{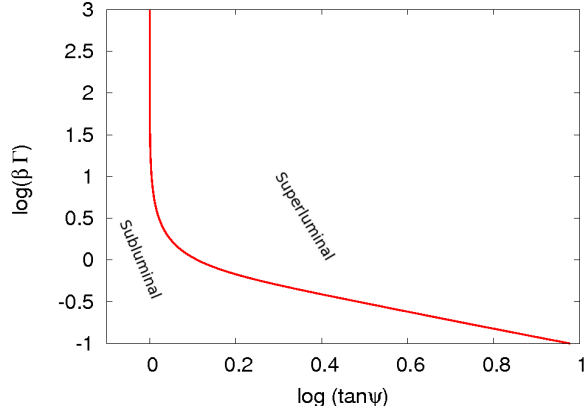}
\caption{\small{Allowed shock inclination angles of a Lorentz transformation into the de Hoffmann-Teller 
frame following $\beta_{HT}\leq \beta_{NSH} \cdot \tan\psi$. The ordinate shows 
the logarithm of the $\beta\Gamma$ product of the intersection point velocity of the flow to the shock surface, while 
the abscissa shows the log($tan\psi$) where $\psi$ is the angle between the flow and the shock normal, in the normal shock frame. 
In the de Hoffmann-Teller frame the vector of the velocity of the flow and the vector of the magnetic field are parallel to each other.}}
\end{center}
\end{figure}

\begin{figure}[t]
\includegraphics [height=5cm, width=8cm] {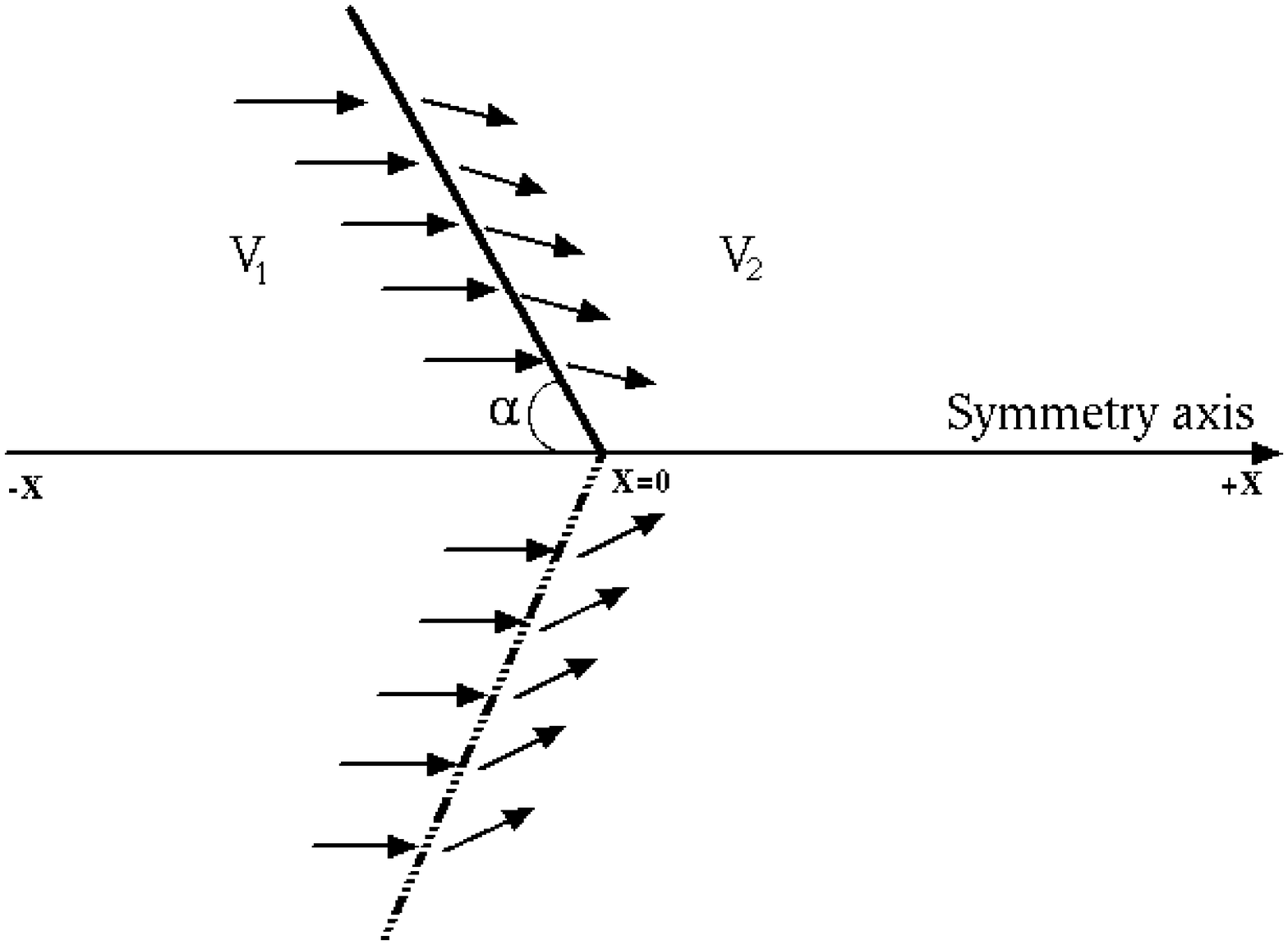}
\includegraphics [height=5cm, width=8cm] {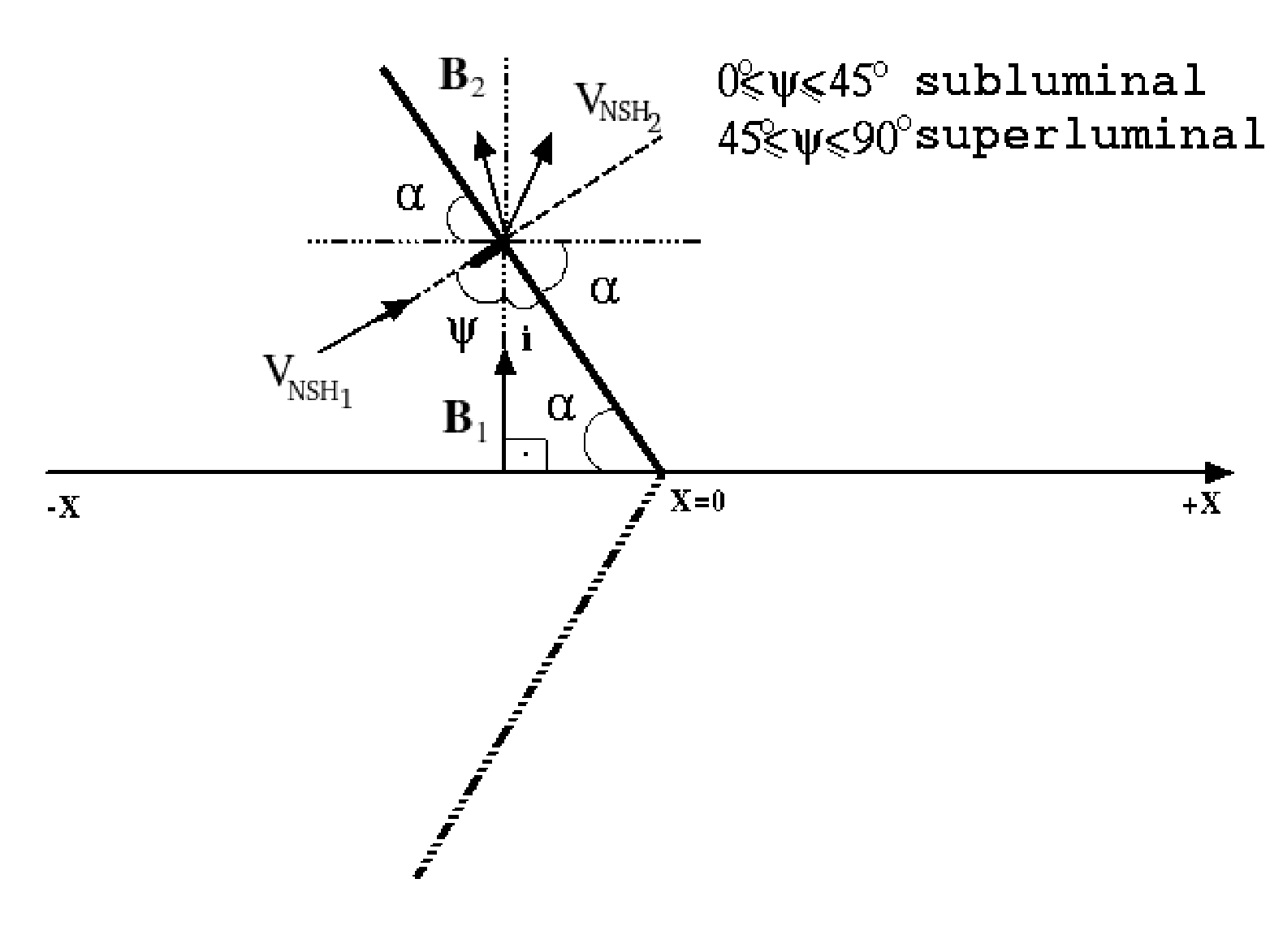}
\begin{center}
\includegraphics [height=5cm, width=8cm] {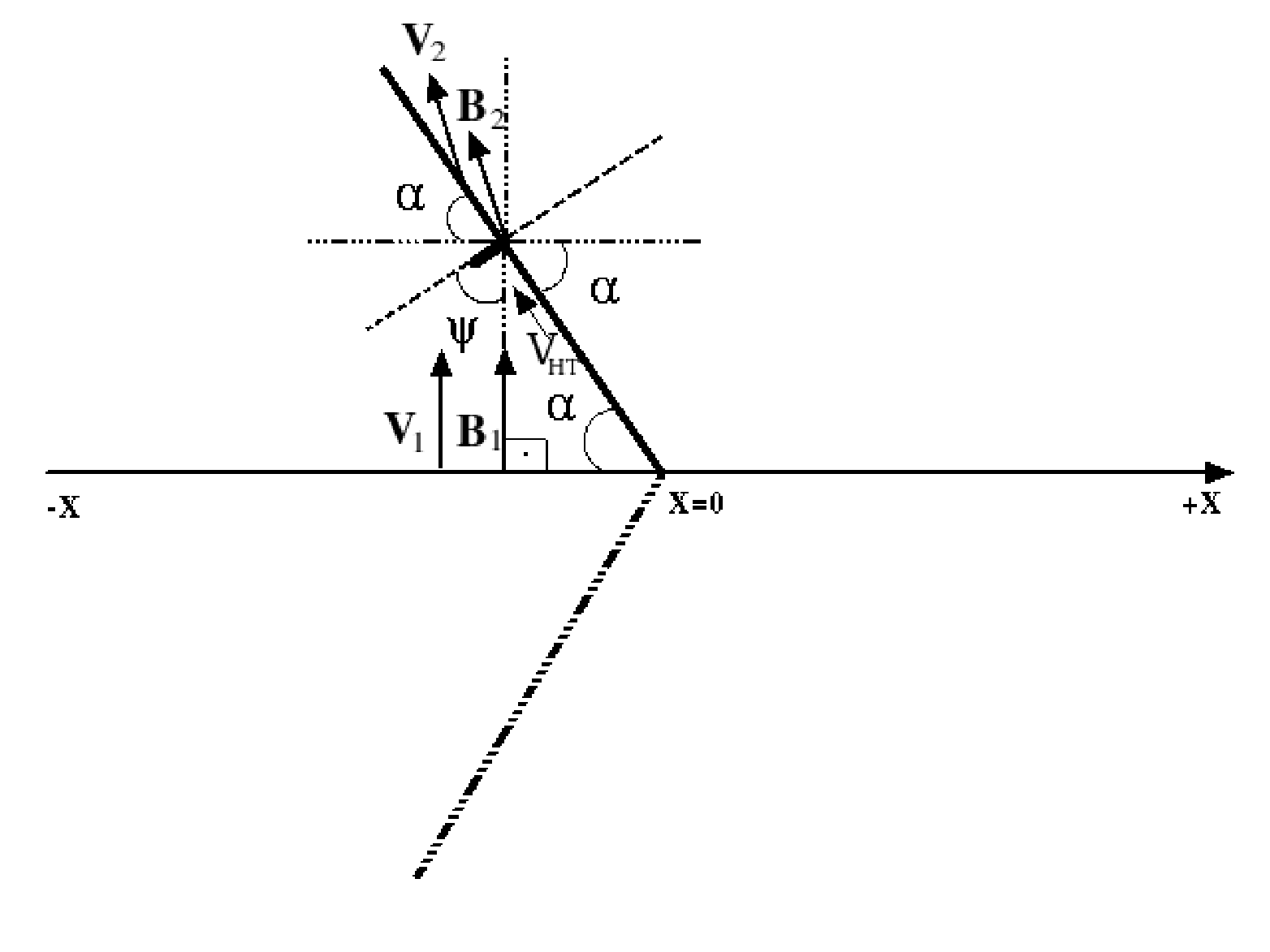}
\caption{\small{Left panel: General topology of the flow in a conical shock.
Right panel: A conical shock as viewed in the normal-shock-frame (NSH),
where the vector of the upstream flow velocity is parallel to the shock normal. Viewing closely the topology in NSH frame, 
one sees that $90^{o} = i + a$ and $90^{o} = \psi + i$ then  $\psi = a$. 
The magnetic field is oblique to the shock surface by definition,  so here, in order to facilitate our view clearer on the geometry of
a conical shock in a local Cartesian system, we assume the case of the magnetic field vector (B), perpendicular to the 
jet axis thus inclined to the shock normal.
Bottom panel: The same conical shock as it viewed after the Lorentz transformation into the
de Hoffmann-Teller frame (HT). In this frame the velocity vector is placed parallel to the vector 
of the magnetic field as was viewed in the NSH frame above. By this transformation there is no electric
field E in this frame.
}}
\end{center}
\end{figure}

\begin{figure}[t]
\includegraphics [height=7cm, width=7cm]{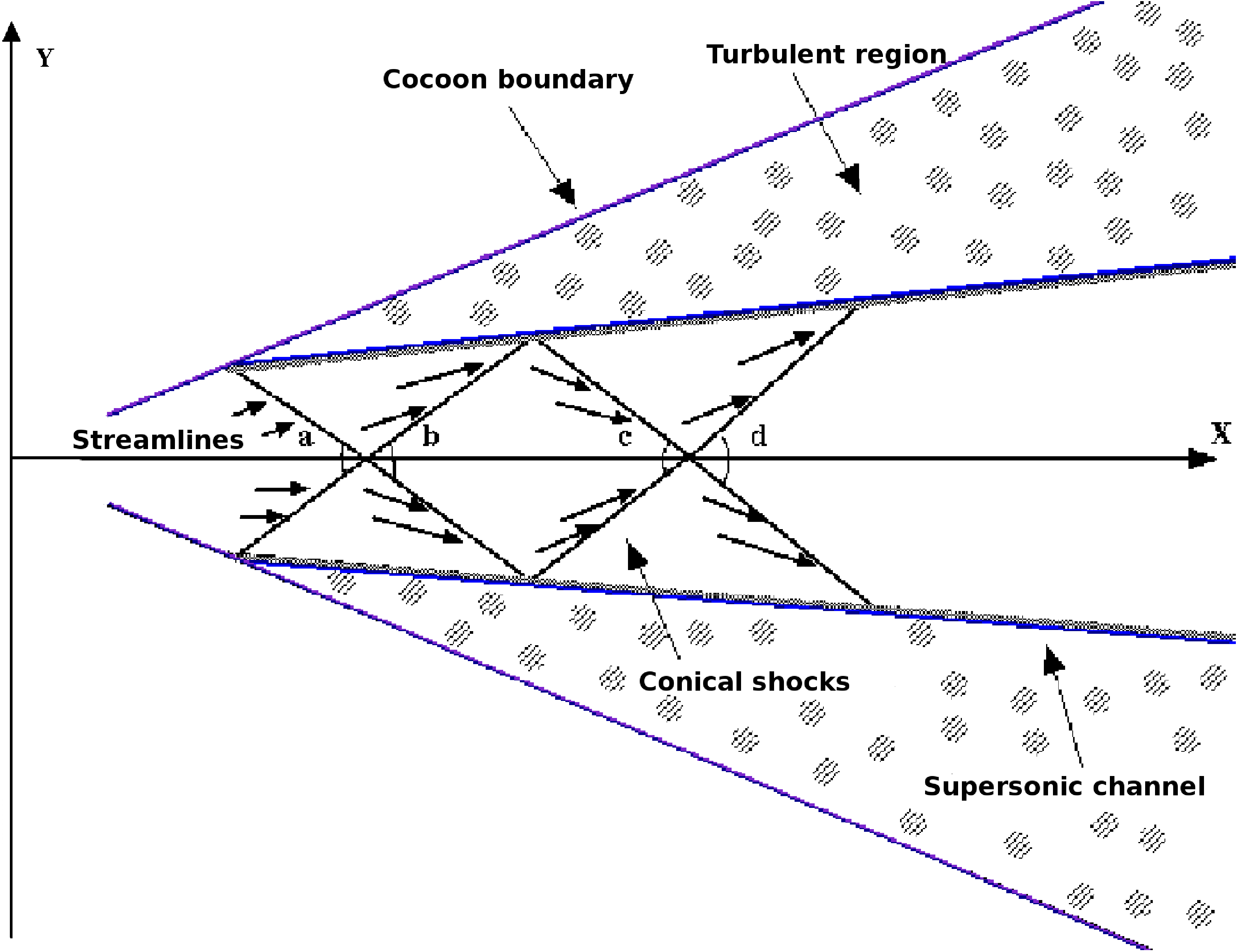}
\includegraphics [height=7cm, width=7cm]{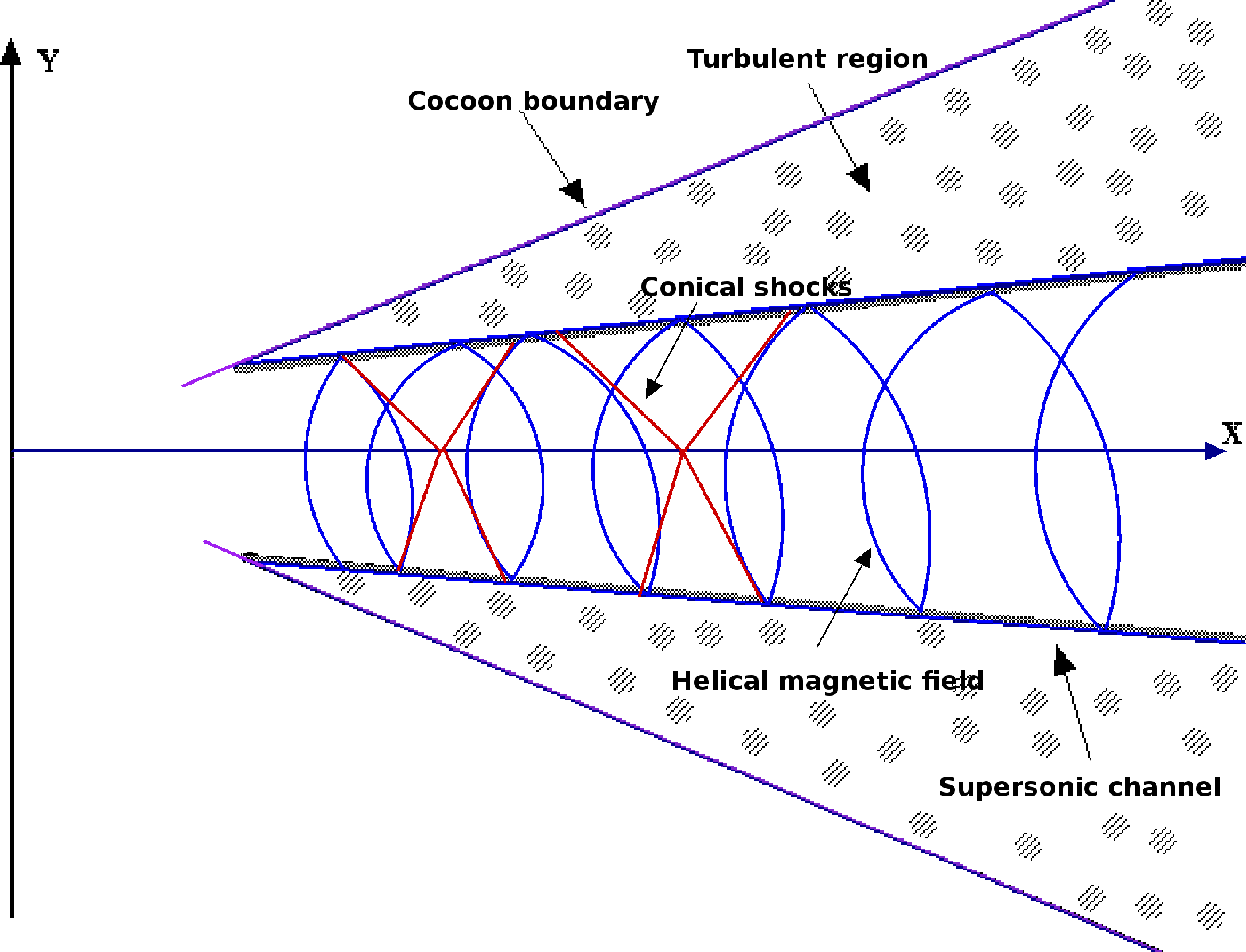}
\caption{\small{The concept of conical shock structures and helical magnetic field topologies: Left, 
repeated conical shock structure of opening angles $a, b, c, d$, in an AGN jet viewed 
at the shock frame. Right, an overall view of the topology of the magnetic field within an AGN jet (not to scale).
}}
\end{figure}


\begin{figure}
\begin{center}
\includegraphics [height=7cm, width=9cm]{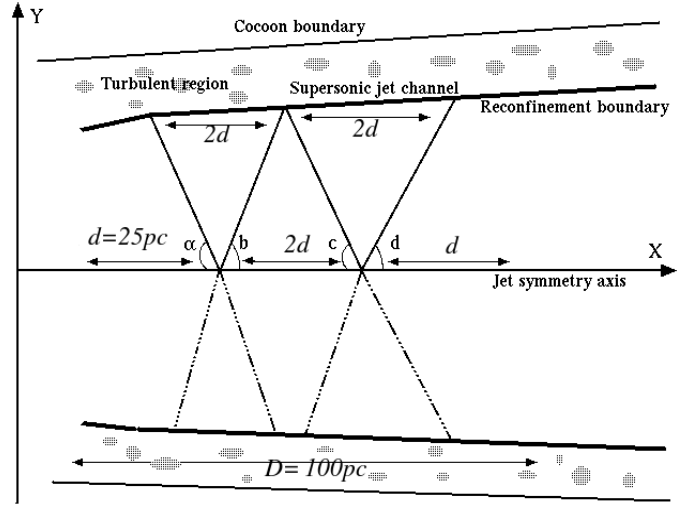}
\caption{\small{An overall view of the proposed jet topology and conical shock geometries (not to scale) simulation framework.
Here $d$ denotes a prefixed distance from the shock, $D$ is the total length of the acceleration area along x-axis, 
which we assume here to be 100 pc.
}}
\end{center}
\end{figure}

\subsection{Individual-oblique shock studies}

Here we will present and discuss past numerical studies for individual relativistic shocks, 
firstly because the Monte Carlo code of Meli and Quenby (2003a,b) for individual shocks is 
extended for \textit{multiple} subluminal and superluminal relativistic shock studies to be 
presented later on, and secondly because we want to consolidate and facilitate the understanding of 
the current results.

First simulation studies were conducted for individual oblique relativistic shocks in Meli and Quenby (2003b). 
Three cases were examined:  1) the dependence of the particle spectra to the scattering model, namely pitch angle diffusion or 
large angle scattering, 2) the effect of relativistic shock speeds to the shape and spectral index of the particle spectrum,
3) the dependence of the produced particle spectra on the magnetic field inclination, as would apply to the superluminal or subluminal 
shock-magnetic field orientation cases.

Specifically, we studied large angle scattering ($1/\Gamma\ll \delta \theta \leq \pi$) and the extreme case
of pitch angle diffusion ($\delta \theta \leq 1/\Gamma$), for a discussion see the Appendix.
The turbulence of the media connected to the chosen value of the mean 
free path was set to $\lambda = 10 r_g$  (see Appendix). 
The velocities of the shock were varied, with  $\Gamma =(1-V_1^2/c^2)^{-0.5}$ values between $5$ and $10^3$, relevant to 
models of ultra-relativistic particle shock acceleration in AGN central engines and relativistic jets and Gamma 
Ray Bursts (GRBs), where $V_1$ is the shocks speed. We moreover considered, 
other than the relativistic value of compression ratio $r=3$, a value of $r=4$ with $\lambda/r_{g}$ larger than 1, 
as a direct comparison to the work of Kirk and Heavens (1989) with similar results.
Those simulation studies have shown that the spectral shape and index depend on whether 
the particle scattering is small angle or large angle, in very good agreement with similar studies of e.g. Baring (2005).
Specifically, for the subluminal (quasi-parallel) shocks the large angle scattering case exhibited distinctive structure 
superimposed on the basic power-law spectrum, largely absent in the pitch angle case. 
A $\Gamma^{2}$ energy enhancement factor in the first shock crossing cycle and a significant energy multiplication 
in the subsequent shock cycles were also observed.
Both scattering operators yielded significant faster acceleration rates when compared with the conventional 
non-relativistic expression for the time constant, or alternatively with the time scale $r_g/c$, an effect found in e.g. Quenby and Lieu, (1989), 
Ellison et al. (1995), etc.
For the (quasi-perpendicular) superluminal shock cases, we examined the energy gains and the spectral shapes. 
Very interestingly, we found that a superluminal (shock-drift) mechanism yields steeper spectra and it is less efficient in 
accelerating particles to the highest energies observed, as it was also shown in Niemec and Ostrowski (2004).

A general trend was that as relativistic the shock speed, as flatter energy distributions exhibited, a 
trend seeing also in e.g in Stecker et al. (2007). As we will see later-on this behavior will be repeated yet again in 
multiple shocks.

Furthermore, in the work of Meli et al. (2008) and Meli (2011) systematic Monte Carlo simulation 
studies were conducted for individual subluminal and superluminal relativistic shocks. 
We used a large range of Lorentz factors, $\Gamma$, and we altered the diffusion scattering compared to the earlier
work of 2003, in order to investigate the effects and differences on the efficiency of acceleration. 
We applied a pitch angle scattering allowing $1 /\Gamma \leq \delta \theta \leq 10/\Gamma$, and decreased the turbulence of the 
media as $\lambda = 10 r_g$,  see also Appendix.
We note that the same  turbulence and diffusion parameters are used for the present simulations. 
It was shown that for very high shock speeds up to a few hundreds of Lorentz factor $\Gamma$,  
subluminal shocks were very efficient in accelerating particles 
up to $10^{12}$~GeV, factors of $10^{9\rightarrow11}$ above the particle injection energy, giving off flat spectra, 
$E^{-\sigma}$, with $\sigma$ values ranging between 1.1 $\pm$ 0.1 and 2.1 $\pm$ 0.1, 
while superluminal shocks were effective only up to $\sim 10^{5}$~GeV, resulting in 
steeper indices of  $\sigma \sim 2.5 \pm 0.1 $. 
\begin{figure}[t!]
\begin{center}
\includegraphics [height=6cm, angle=270, width=6cm]{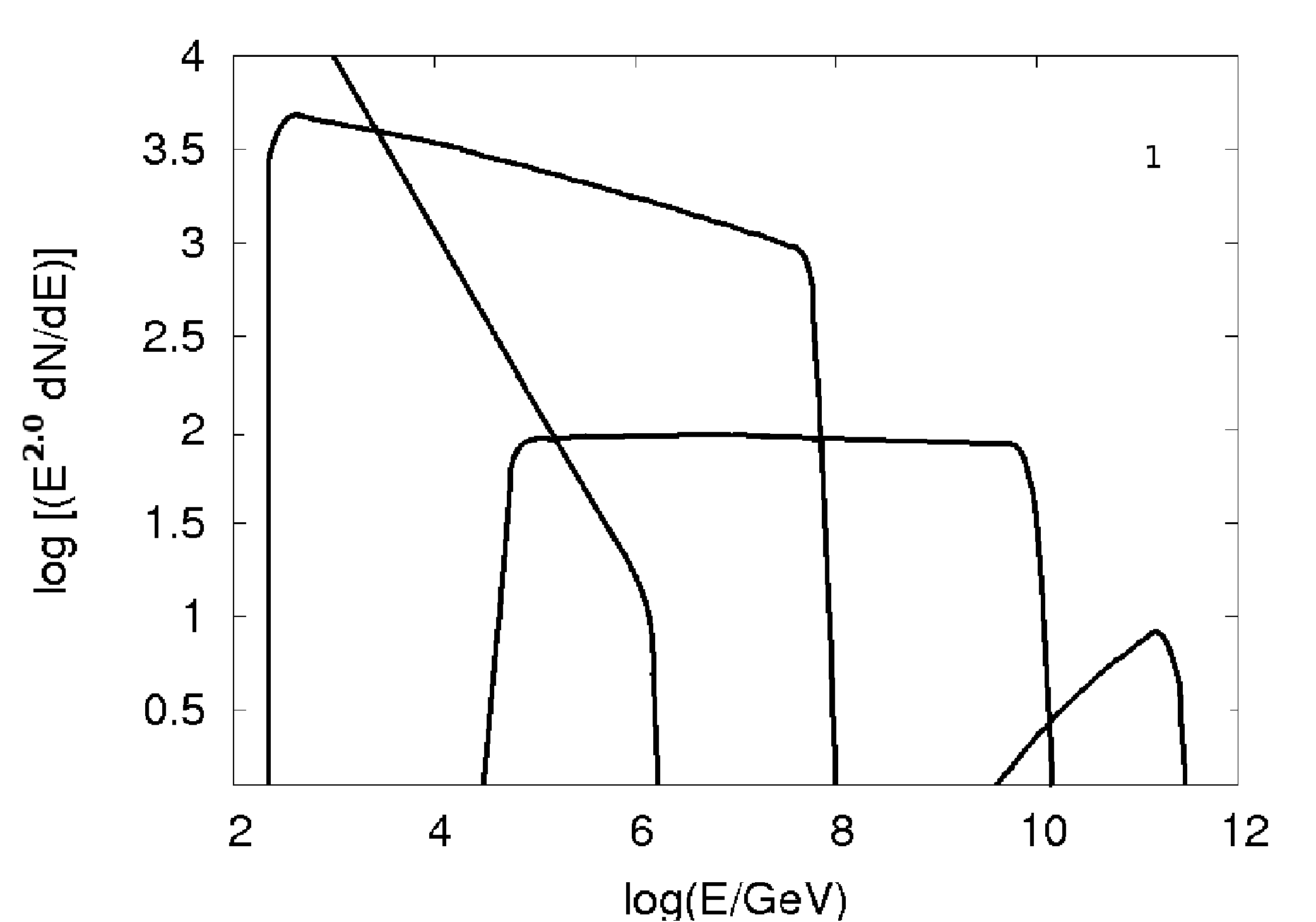}
\includegraphics [height=6cm, angle=270, width=6cm]{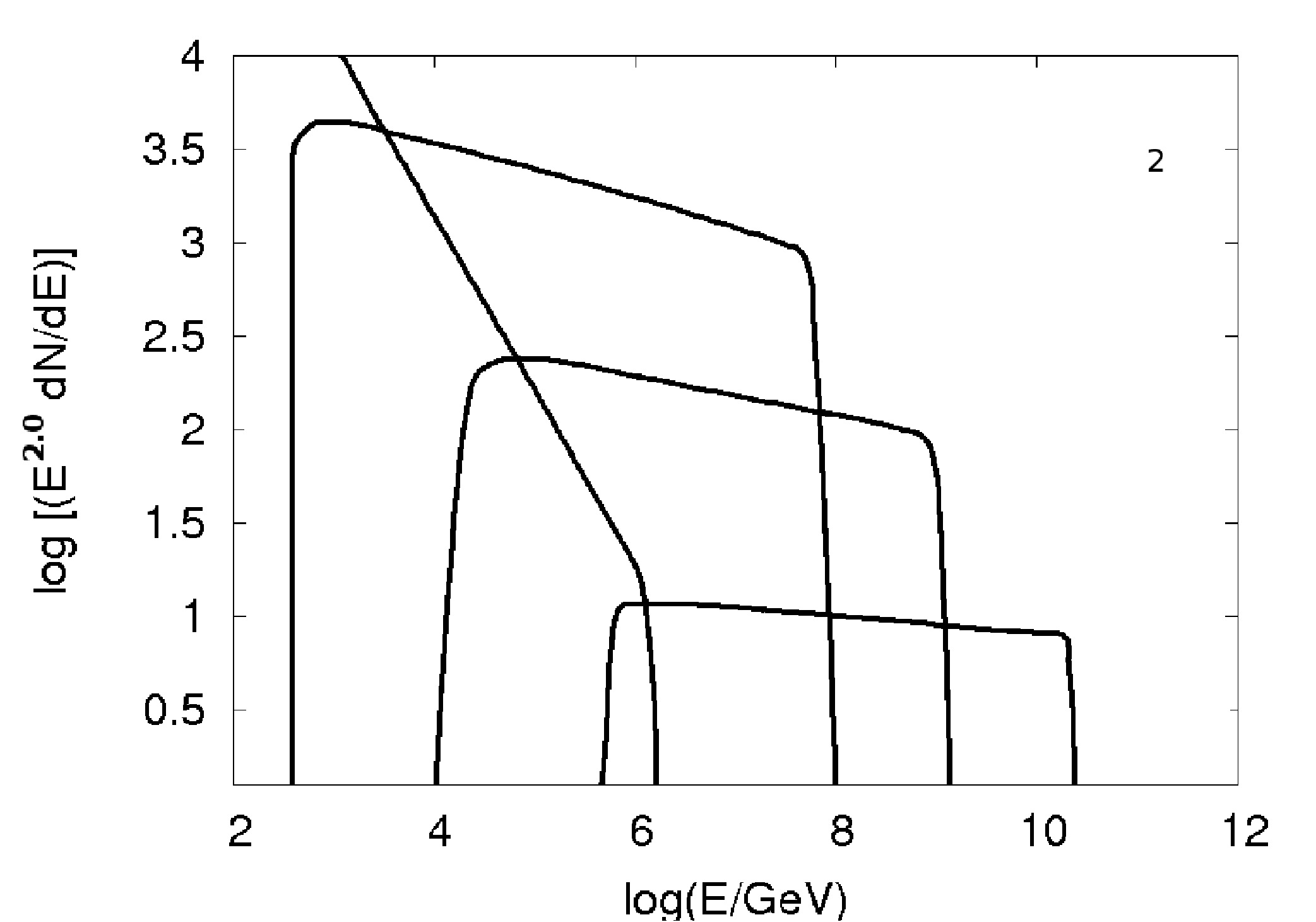}
\includegraphics [height=6cm, angle=270, width=6cm]{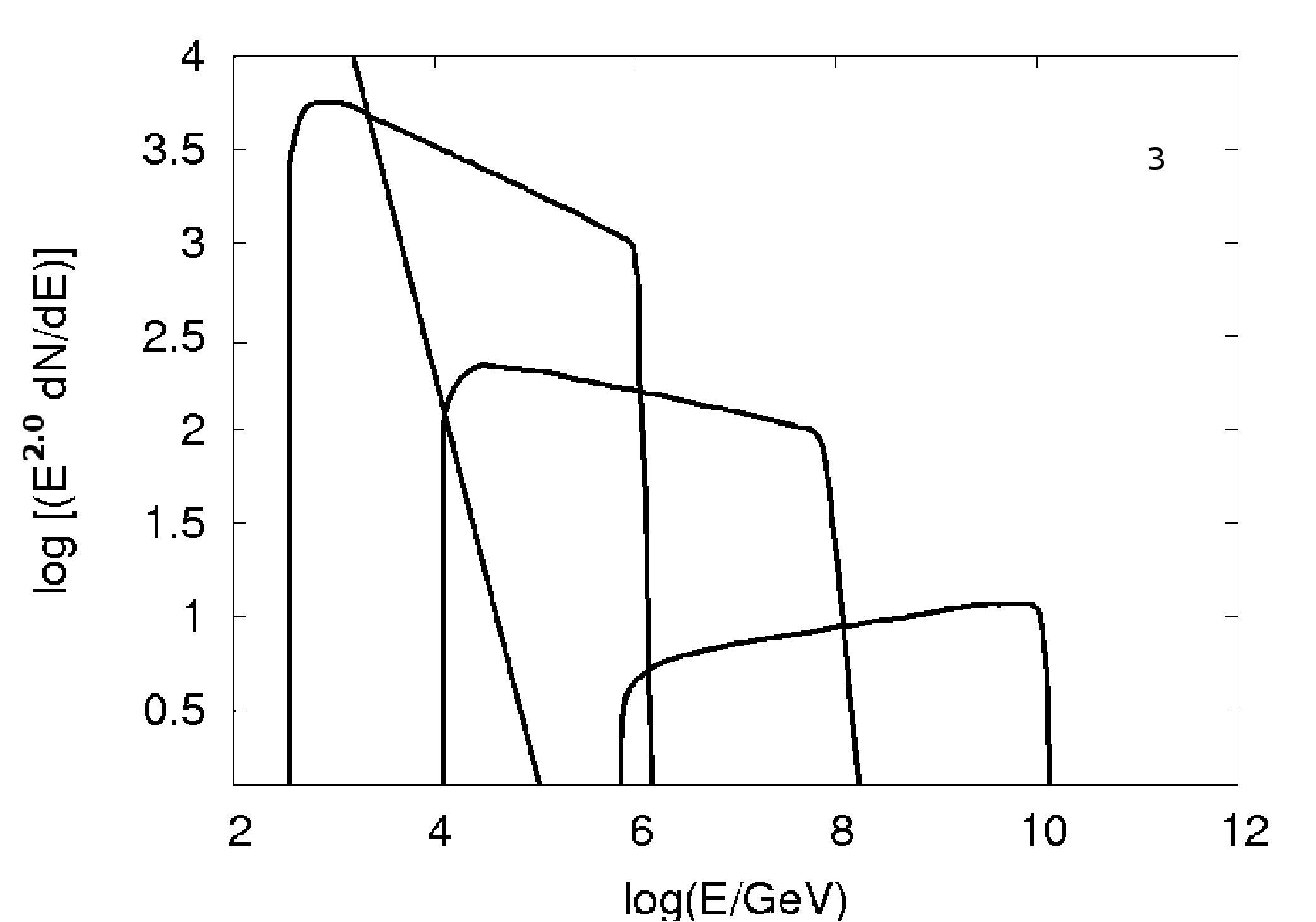}
\includegraphics [height=6cm, angle=270, width=6cm]{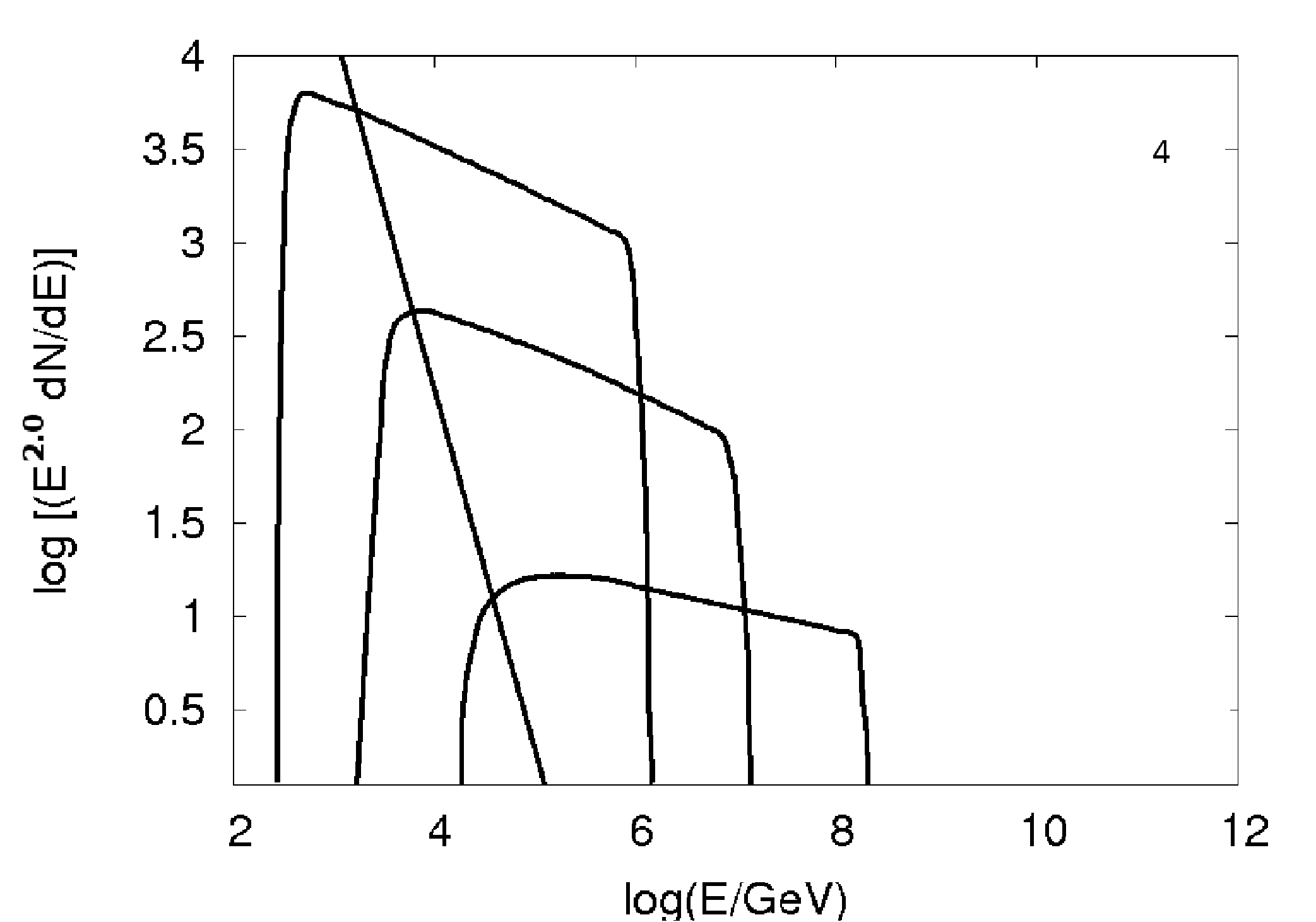}
\includegraphics [height=6cm, angle=270, width=6cm]{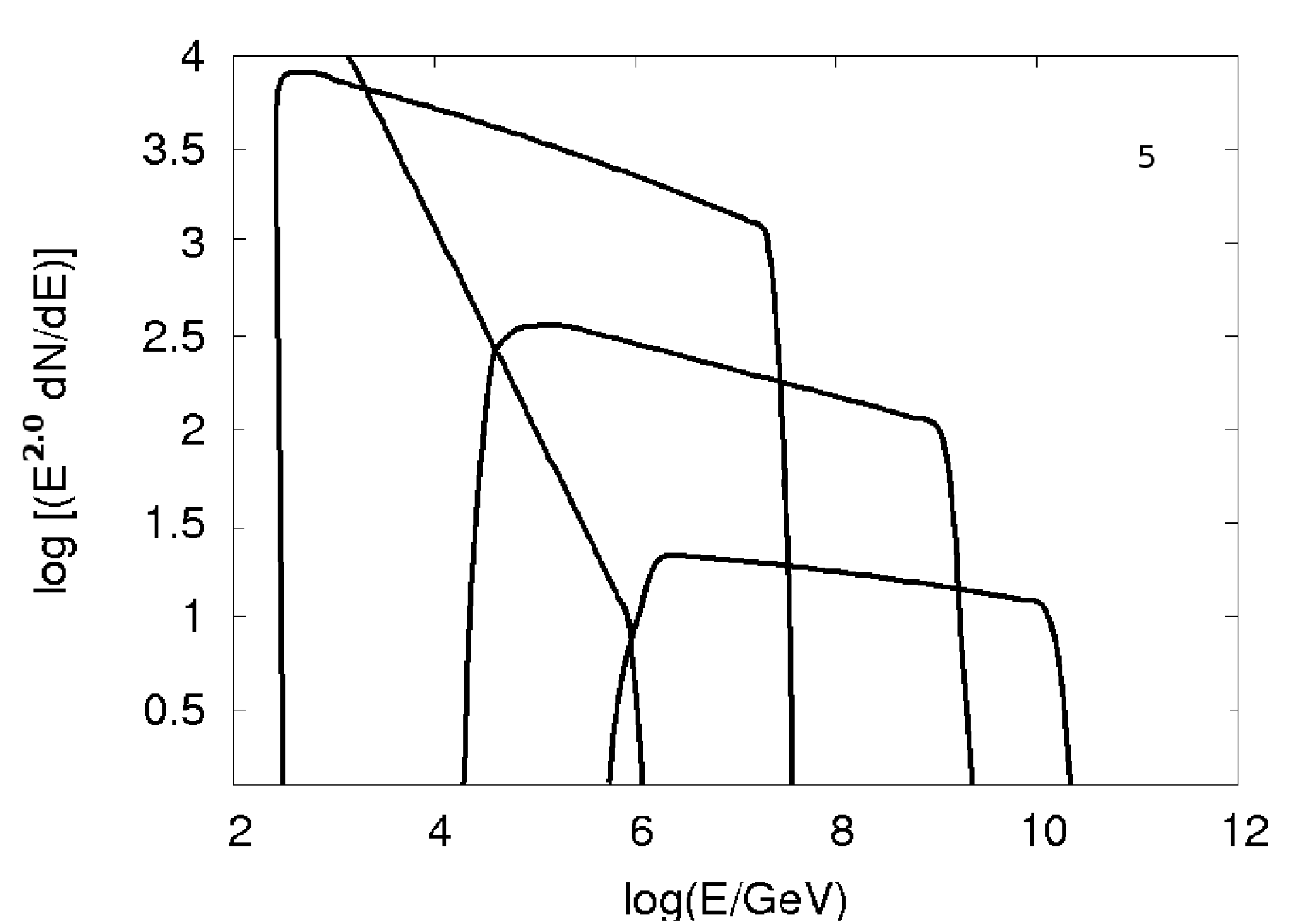}
\includegraphics [height=6cm, angle=270, width=6cm]{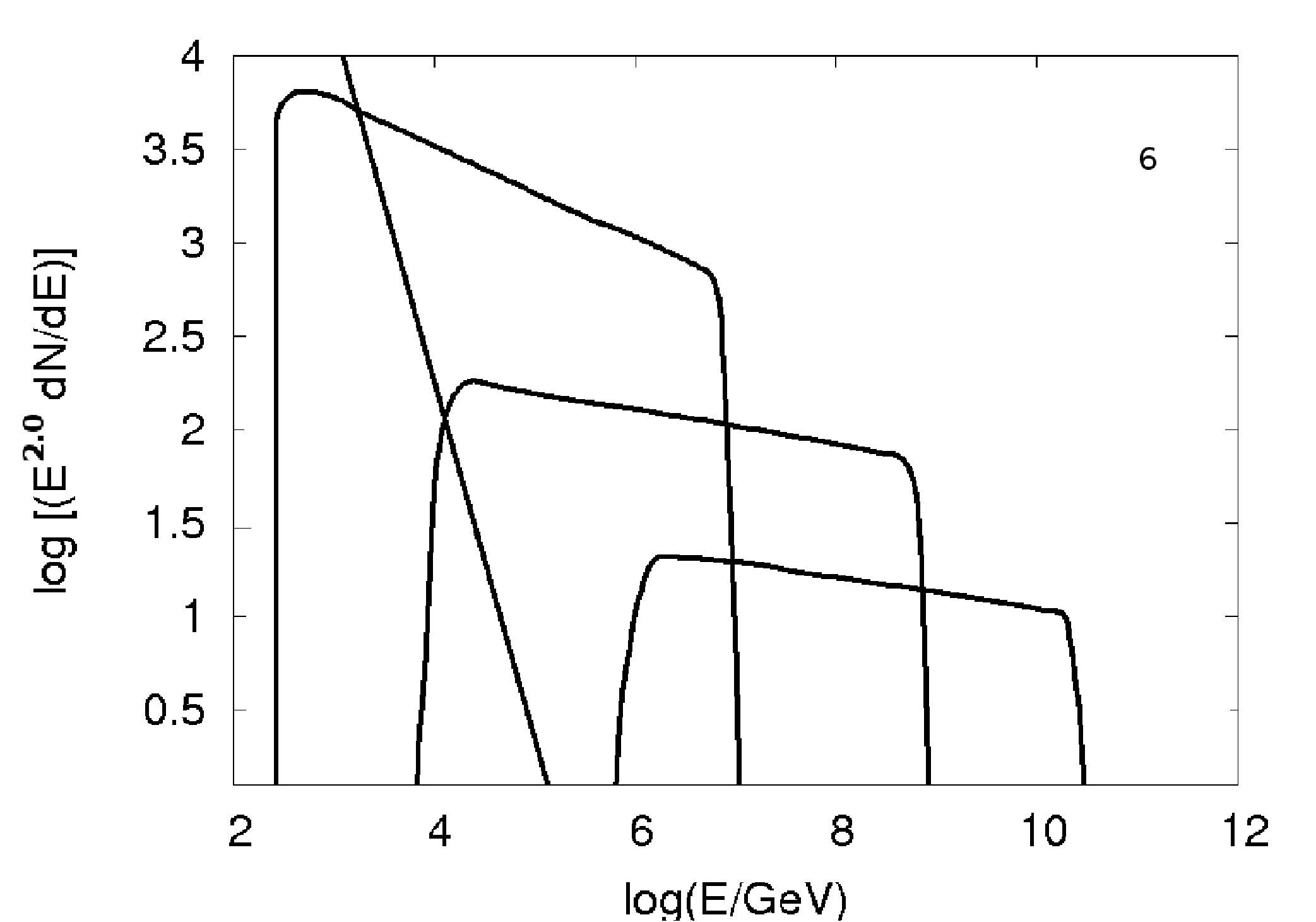}
\caption{\small{
Differential spectra (multiplied by $E^{2})$, calculated in the respective shock frame downstream, by sets of four oblique consecutive relativistic shocks. 
Here one sees developed flatter spectra (from left to right respectively) extending to very high energies with a depletion 
on the lower energies (analogy to figure 1).
Panel 1: Spectra by four consecutive subluminal shocks of inclination $\psi=45^{o}$ (corresponding to a half opening angle).
Panel 2: Spectra from two subluminal shocks and two superluminal shocks 
($\psi=85^{o}$). 
Panel 3: Spectra by two superluminal and two subluminal shocks.
Panel 4: Spectra by four superluminal shocks all four for the same inclination $\psi=85^{o}$.
Panel 5: Spectra by a sequence of subluminal-superluminal-subluminal-superluminal shocks.
Panel 6: Spectra by a sequence of superluminal-subluminal-superluminal-subluminal shocks. See complimentary Tables 1-6.
Spectra} have been shifted vertically to allow for better comparison.}
\end{center}
\end{figure}

These lines of approach that led to these results are justified by observational evidence e.g. Dingus et al. (1995), 
regarding irregular and flat spectra from GRBs
as well as many flat or irregular spectra observed by Fermi telescope, etc. The results were also consistent with observations 
of the electron spectra that may be injected from 
the terminal hotspots and lobes of the powerful FR-II radio galaxies which are not of a single and universal power-law 
form, as shown in detail in Rudnick et al. (1994), Machalski et al. (2007), and others.

These numerical studies  about individual shock acceleration (which are the basis of the present study)
have shown an absent 'universal' spectrum,  
with a variety of spectral indices depending on the scattering model, shock inclinations and 
speeds, with an overall noticeable efficiency and flatness developing at higher $\Gamma$. 
Different diffusion media gave rise to a multitude of spectral 
slopes  as well as some structure in the developed particle distributions, mostly evident for 
large angle scattering.  On the other hand different shock inclinations (sub- or super-luminal shocks) 
affect the acceleration efficiency and spectral slopes (for a detailed discussion see also 
Meli, 2011).
The above findings are in excellent agreement with similar works of 
Ellison and Double (2004), Niemiec and Ostrowski (2004), Stecker et al. 
(2007) and Baring (2011) and all compare well with each other in terms of their 
predictions and understanding.
As we will see in the following section, the properties of individual relativistic shocks are carried on into
the multiple shocks, with some additional properties that we will describe below.

\subsection{Multiple-oblique shocks}

In the non-relativistic limit authors such as e.g. White 1985, Melrose \& Pope 1993, Schneider 1993, Pope \& Melrose 1994, etc, 
solved the problem of subsequent multiple shocks analytically, calculating the spectrum as it is processed through a number of shocks, 
suppressing new injection, and considering adiabatic decompression.
It has been analytically shown that the downstream distribution of particles
after acceleration in a number of shocks with decompression between them, establishes a 
power law, with flatter distributions comparing to a single shock acceleration e.g. Melrose \& Pope (1993) and Gieseler and Jones (2000).
In the limit of an infinite number of subsequent shocks with
injection at each shock, the flattening of the spectrum (compared to a single shock) extends even to the highest energy particles,
with a momentum dependence of $f(p) \propto p^{-3}$ (White 1985, Achterberg 1990). It is interesting to note here that the above behaviour 
seems equivalent to a shock of infinite density jump (Drury, 1983).

Here we will investigate acceleration of particles in relativistic superluminal and subluminal multiple shocks. Given blobs of ejected 
hot plasma by the black hole of an AGN can travel along its jet and pass through  a sequence of shocks,
as it has observed for example in the case of BLac PKS 1510-089 (Marscher et al. 2010),
we allow a single initial injection of particles at the beginning of the acceleration 
of a fixed number of particles ($N_i=10^4$ of an initial $\gamma=\Gamma_{sh}+10$) and
we use an initial shock Lorentz factor ($\Gamma = 50$) for the first conical pattern (which consists of two oblique shocks 
with the same apex and inclination angles $a$ and $b$ to the jet axis ($x$), see figure 4). 
For the second pair of shocks (with inclination angles $c$ and $d$) we use a $\Gamma = 17$, and this could be physically 
justifiable due to the decompression-compression of the downstream plasma
of the precedent shock pattern (i.e. velocity compression ratio of 3).

As the particle properties are measured in the local plasma frame according to the relativistic jump conditions, 
the downstream properties for a shock become upstream properties for the subsequent one at a fixed  $d = 25pc$, see figure 5.
We note that the shocks are simulated within a single simulation box, so some particles 
travel far upstream to a precedent shock and continue  the acceleration. Nevertheless, due to 'beaming' effect
of the particles in relativistic media (Achterberg et al. 2000, Meli and Quenby 2003b, Meli et al. 2008) 
only few make it upstream to a previous shock. 

As aforementioned, during the acceleration process a fixed probability of escape $P_e$ between shocks due to 
decompression effects is also allowed (Melrose and Pope 1993, Pope and Melrose 1994), since one would consider that every particle 
accelerating from one shock to another, will always reach the next one further accelerating. 
It is expected that particles found downstream the 4-th shock of our assumed sequence, 
may have been accelerated in less than 4 shocks, thus a fixed escape probability $P_{e}=0.3$ is allowed
to be applied between the shocks during the acceleration process.
In other words, the escape probability gives the fraction of particles of the downstream
distribution of each shock that will not be further accelerated.
These particles remain in the system and contribute directly to the downstream distribution
of the 4-th shock.


As discussed in section 1.1, theoretically, the shock inclination values may lie 
between perpendicular (superluminal), where a shock-drift acceleration mechanism occurs,  
and oblique (subluminal) ones, where diffusive acceleration takes place in the HT frame.
One can see that in the relativistic case half of the inclinations allowed are superluminal 
and the rest are subluminal,  due to the $45^{o}$ transformation frame inclination limit, 
see  figure 2 and section 2.1.

It is shown by observations (e.g. Hughes et al. 1985, Cawthorne \& Wardle 1988) 
that when the jet contains an initially tangled magnetic field, 
plane shocks perpendicular to the jet axis produce polarization with an 
electric field perpendicular to the shock front (parallel to the shock normal), 
and hence parallel to the jet direction. 
Observationally, given a parallel electric field to the jet axis it means that the 
magnetic field is perpendicular to the shock normal, therefore particles trapped
into the frozen-in plasma, will get accelerated by repeatedly crossing the shock plane.
One sees that this kind of topology
serves as a large scale analogon to the non-diffusive shock-drift acceleration mechanism we aforementioned, 
where a superluminal case applies.
Nevertheless, as it is discussed in Section 1.1 theoretical and numerical studies show that the inclination 
of the oblique shocks inside a supersonic jet channel must have small angle values to the jet axis in order 
for a strong acceleration effect to occur. Therefore we assume that in the same manner, identical superluminal shocks are 
rather unrealistic to occur during the acceleration phase of the AGN jet.

Here we assume protons with no losses, as we are interested in the produced spectra at source,  which can be
justifiable since the magnetic field in AGN jets can be weak (i.e. B =$10^{-4}$ Gauss).
Moreover, the temperature of an ambient photon field in a jet can be low, which imposes 
only a weak loss factor for protons, applying in the considered scenario. In that sense the initiation of 
secondary cascading e.g. $p\gamma$ interaction, is minimal. This is  also in accordance with 
field conditions far off the accretion disk-jet interaction area of the AGN ($> 3000$ Schwarzschild-radii), 
see  Markoff et al. (2005), Becker and Biermann (2009).

Here we perform a series of simulations for a range of shock Lorentz boost speed $\Gamma$ ($10 < \Gamma <50$),  
and we find quite similar spectral behaviors.
Therefore, the present Lorentz shock speed chosen is an exemplary case (as the velocity of plasma in AGN jets 
typically ranges between around 10 and 50 Lorentz factors).
Six different sets of four consecutive oblique shocks are simulated for comparison purposes,  and their differential 
spectra are shown in figure 6. The spectra are recorded in the shock-rest frame downstream each shock, at
a distance $d$, see figure 5.
We note that we allow two shock sets with identical inclination angles between them, for simplicity and for
direct comparisons, and because physically a decompression between shocks (e.g. Melrose $\&$ Pope 1993) 
can restore the magnetic field to the initial upstream shock value and orientation. 
Nevertheless, there could be many other situations of non-identical inclinations between subsequent shocks
therefore, thus we allowed this condition  other shock set profiles as well.

\begin{figure}[h!]
\begin{center}
\includegraphics[height=1.8cm, width=5cm]{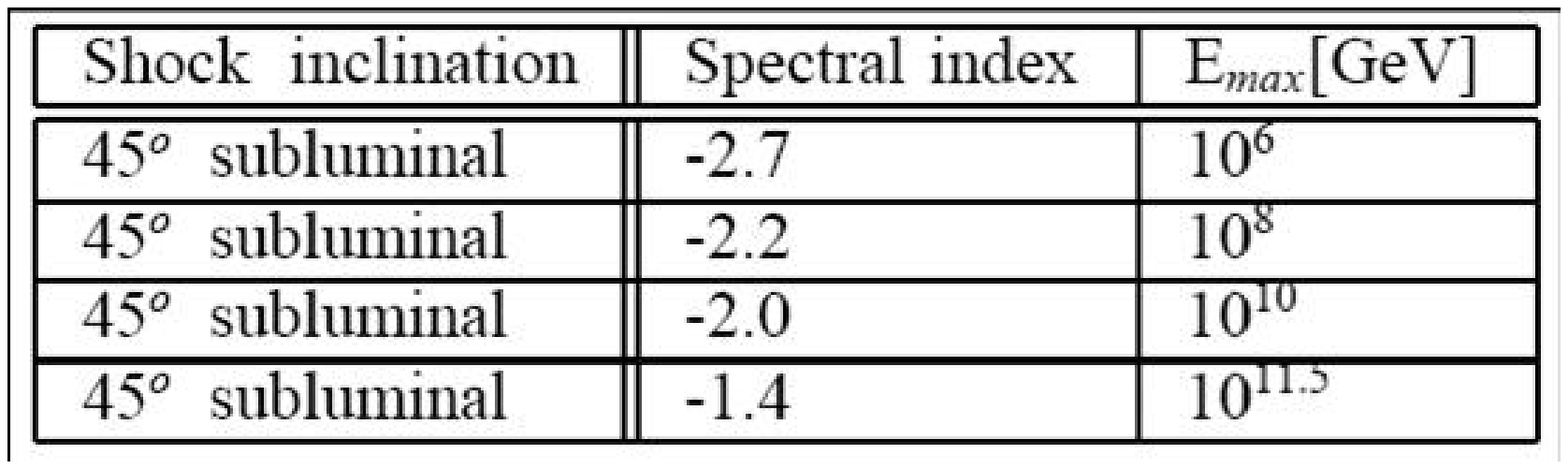}
\includegraphics[height=1.8cm, width=5cm]{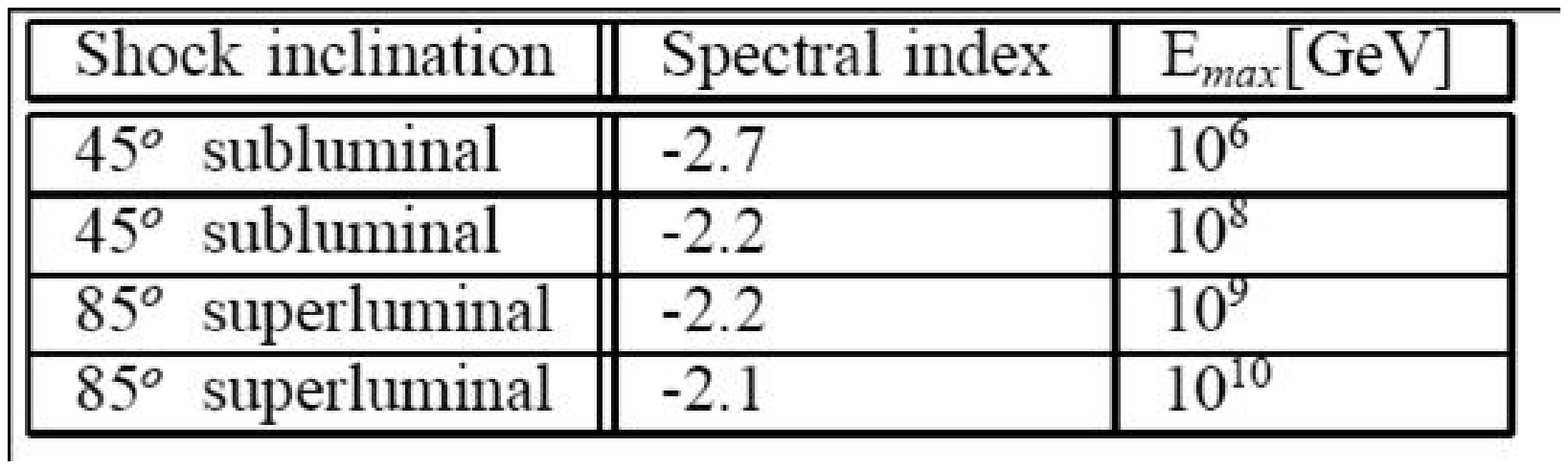}
\includegraphics[height=1.8cm, width=5cm]{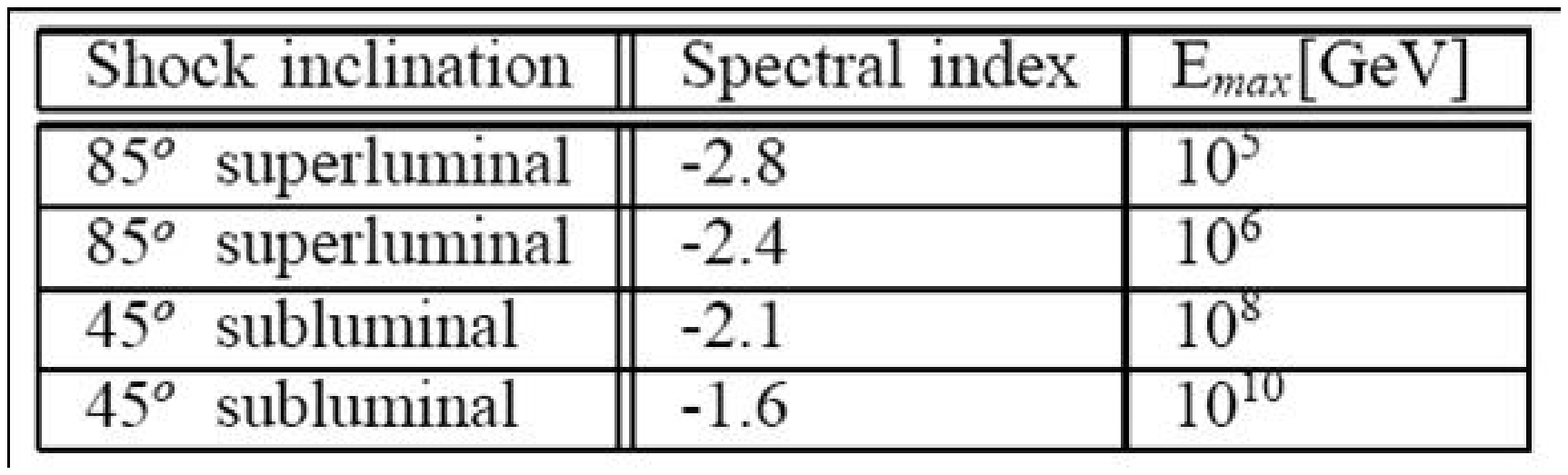}
\end{center}
\end{figure}

\begin{figure}[h!]
\begin{center}
\includegraphics [height=1.8cm, width=5cm]{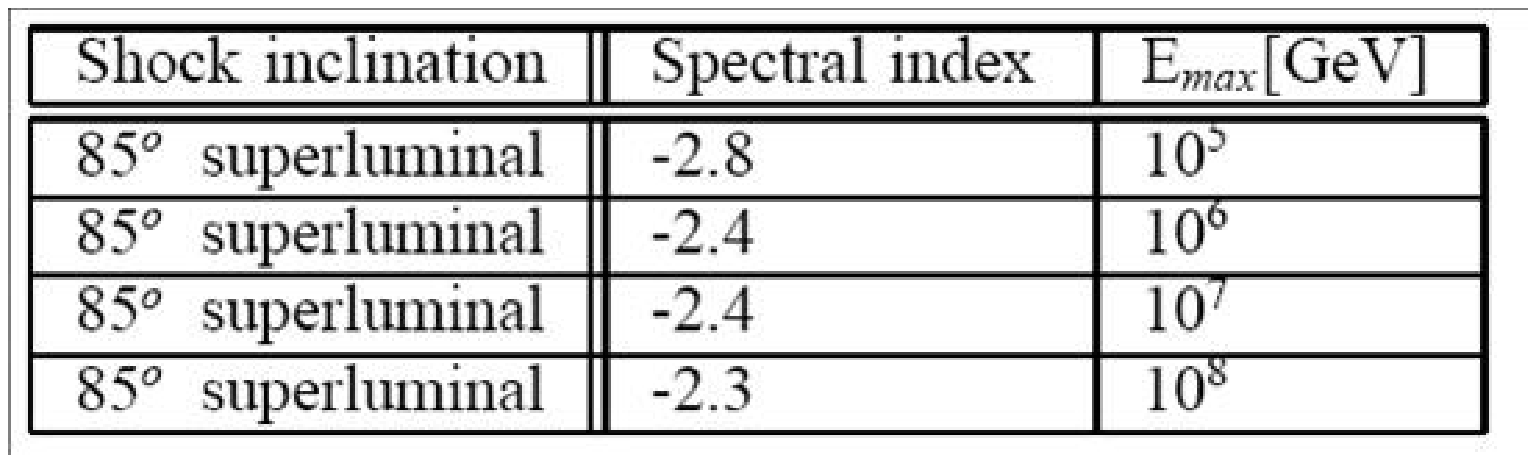}
\includegraphics [height=1.8cm, width=5cm]{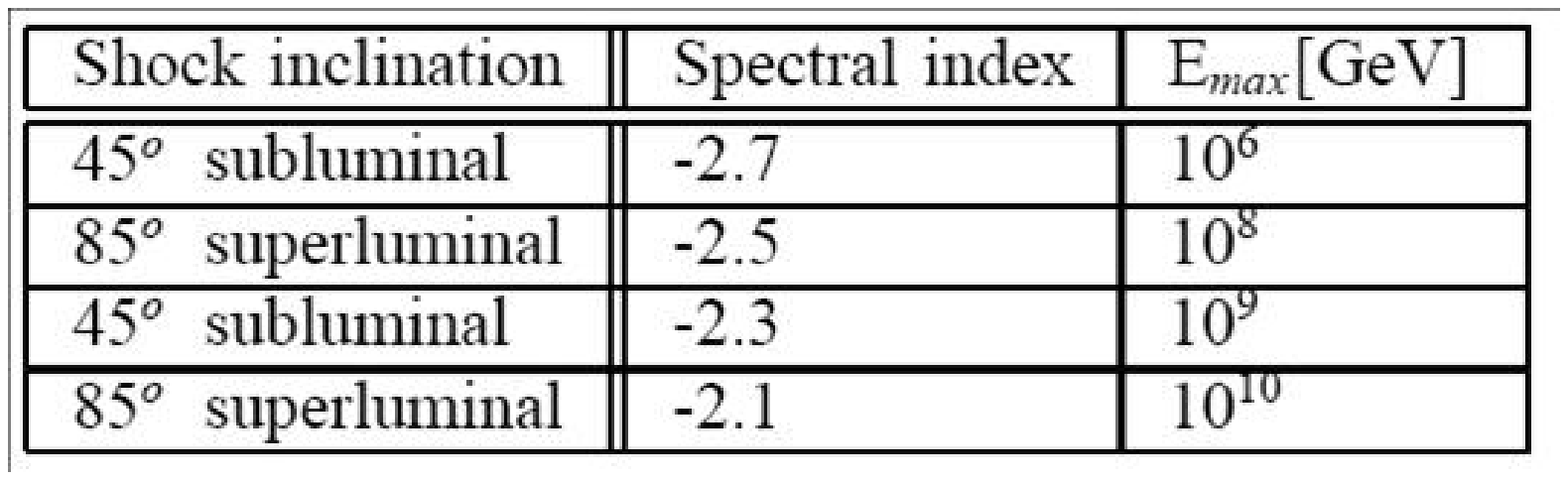}
\includegraphics [height=1.8cm, width=5cm]{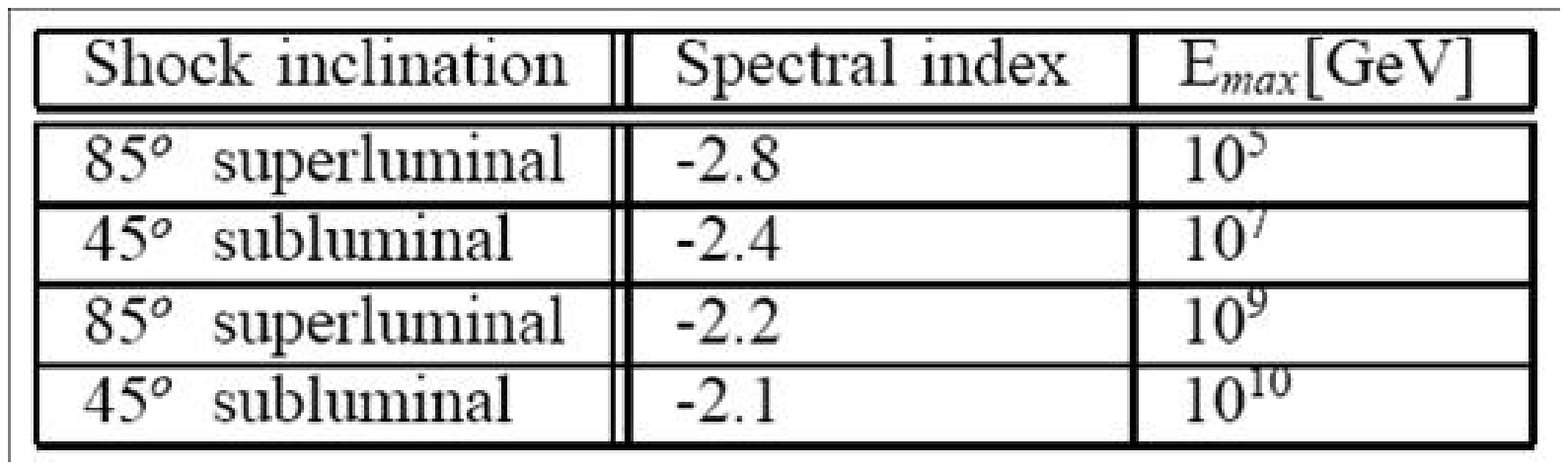}
\end{center}

\begin{center}
\small{{\bf Tables 1-6}. Shock inclinations, spectral indices and maximum energies attained for the six cases of four 
shock combinations (complimentary to the spectra shown in figure 6). Each table corresponds to each panel of figure 6 (left to right).}
\end{center}
\end{figure}

In order to facilitate our understanding, in figure 6 we depict the respective differential particle spectra 
by each shock in the same plot. The spectra have been shifted vertically to allow for better comparison.
Specifically, in panel 1 we show spectra by four identical consecutive
subluminal shocks,  each with a half opening angle ($a, b, c,  =45^{o}$) 
to the jet flow, equal to the actual inclination of the shock to the magnetic field as we noted in paragraph 3.1, see figure 3.
Panel 2, shows spectra for two subluminal ($a, b =45^{o}$) and two superluminal shocks ($c, d=85^{o}$).
Panel 3, indicates spectra for two superluminal and two subluminal shocks while panel 4 shows
spectra from four identical superluminal shocks with the same inclination angles ($a, b, c, d =85^{o}$). Plots 5 and 6  
show a set of subluminal-superluminal-subluminal-superluminal shocks and superluminal-subluminal-superluminal-subluminal 
ones, respectively. Tables 1-6  are complimentary to figure 6, showing values for the inclination 
of the shocks in each set, the spectral indices and the maximum attained energies by each shock.
Since we are dealing with relativistic shocks, 
as aforementioned, the pitch angle scattering is chosen as $1 /\Gamma \leq \delta \theta \leq 10/\Gamma$, with media turbulence as 
$\lambda = 10 r_g$,  based on the analytical description given in the Appendix.

Firstly, by inspecting panel 1 (figure 6), one sees the maximum particle energy reached of
an impressive $\sim 10^{11}$ GeV. In panel 4, the case of four consecutive superluminal shocks is shown which is the least 
efficient, with a maximum attained particle energy of a modest $\sim 10^{8}$ GeV. 
Physically, the case of four identical superluminal shocks would actually require a repetitive set of oblique shocks with very large 
opening angles ($a,b,c,d \approx 90^{o}$), meaning for a distance of about 100 pc a continuous decelerating flow, 
but as we discussed in section 1.1, this is not realistic.

Secondly, one notices that (i) the spectra by each shock in every shock-set, get gradually flatter, following on one hand 
the findings of individual relativistic shocks (e.g. Ellison et al. 1995, 
Stecker, 2007, Meli et al. 2008), and on the other the trends of White (1985), Melrose and Pope (1993), Schneider (1993), 
Gieseler and Jones (2000), etc, for multiple shocks. Specifically the later work noted that the flat particle 
spectra from multiple shocks is due to the obliquity (identical or not) of the shocks, a result that is in accordance with the present study.
(ii) The flatness of the spectra is more evident for the cases of consecutive subluminal shocks
than superluminal ones, following, as expected, the individual shock effects we discussed in section 2.2. 
It is important to note that superluminal shocks result in steeper spectra compared to subluminal ones, 
but both (relativistic, individual or multiple) result in flatter energy distributions than their non-relativistic counterparts.

Since a fixed number of particles was injected once and
accelerated through more than one shock, this in other words means a gradual 
decreased flux at lower energies and a consequent 'extension' of the energy distributions to higher energies.
For example, in the case of the flattest spectra occurred (panel 1, figure 6)
one sees a change in spectral fluxes as large as six orders of magnitude. The same trend is seen in all other cases.
This is an important result, in terms of source power interpretation, which we will discuss later-on.


In more detail:

\begin{enumerate}

\item The first two shocks in all shock sequences range between spectral index ($\sigma$) values of -2.7 to -2.2 
for subluminal shocks, and -2.8 to -2.4 for superluminal ones.

\item The sequence of four identical subluminal shocks generates the flattest spectra comparing to all other cases, with the
flattest value reaching a $\sigma$ at -1.4, and the highest achieved particle energy at $\sim 10^{11}$ GeV.

\item Four consecutive identical superluminal shocks generate the steepest spectra with $\sigma \sim -2.3$, 
and the lowest maximum particle energy attained of $10^8$ GeV.

\item In all spectra, one sees a flux depletion in lower energies and an additional gradual flatness 
of the slopes, with an extension of the spectra to high energies.  
This is due to the fact that fewer and fewer particles participate in the acceleration at the highest of energies, but this is
balanced by the properties of relativistic shock acceleration, such as acceleration "speed-up" 
(see Meli \& Quenby 2003b) and relativistic subluminal shock acceleration efficiency. 
We note here that our spectra look very similar to those of Melrose and Crouch (1997),
although the latter treated non-relativistic shocks.

\item In all six shock sequences, two consecutive subluminal shocks flatten the spectra by $0.5$. On the other hand, 
their superluminal counterparts remain steeper with only a small spectral index deviation value of $0.1$.
In the first case (see table 1) of the four identical subluminal shocks the spectra flatten by 1.3  with a similar behavior for  
case 3 (see table 3) of two superluminal and two subluminal shocks, with an overall flattening of 1.2.
For the rest of the cases we see an overall flattening by 0.6, 0.5, 0.6 and 0.7 respectively for cases 2, 4, 5 and 6.
Given that the pitch angle diffusion scatter is fixed, the variations of the slopes are connected to the 
individual subluminal or superluminal shock properties discussed in section 2.2, but also on the 
conditions prescribed for multiple acceleration i.e., escape probability $P_{e}$ factor 
between consecutive shocks, identical inclinations of shocks to the jet flow i.e., superluminal or subluminal shocks, or pair 
alteration between the two inclinations, etc.

\end{enumerate}

Examining our model results in an astrophysical context, one derives that the first shock from a repetitive sequence 
in an AGN jet, could establish a low energy power-law spectrum with a spectral index of $\sim 2.7$, while the following shocks 
of the sequence push the particle spectrum up in energy. One observes gradual flatter energy distributions but most importantly the 
flux depletion at lower energies, forming characteristic starved spectra. This means that with a single injection of low energy 
particles through a set of oblique shocks, one can achieve very high energies through acceleration, 
in a set of separate starved spectra but which over a superimposition could give flatter
spectra in very high energies, solving in that way the puzzling apparent source power problem of AGN fed into 
the very high energy cosmic rays.


As we aforementioned in section 1, for particle energies above $\sim 10^{9}$ GeV,  
the best fit injection power-law index value used in the models of de Marco \& Stanev (2005) and Berezinsky et al.
(2006, 2009) is  between -2.4 and -2.7. Specifically, in that work theoretical comparisons were made with observation data
by HiRes  (HiRes Collaboration 2009) and Auger experiments (Auger Collaboration 2007, 2010a, 2010b). Their model assumed a number 
of sources homogeneously distributed throughout the universe. The accelerated particles by these sources were then let to propagate. 
The  initial spectral index necessary to reproduce the observations, assuming that the particles were protons, 
was found to be within the values of -2.4 to -2.7, which is quite close to many cases shown 
in this work. 

In our work it is also shown that consecutive subluminal shocks result in very flat starved spectra, up to UHECR values.
In order to fit the required spectra of de Marco \& Stanev (2005) and Berezinsky et al.
(2006, 2009), one would think that assuming protons accelerated at the source, then 
there must be nearby some dense photon field where over proton synchrotron losses, one could achieve steeper spectra,
but on the other hand this would compromise the energies reached in the primary accelerated population. Thus it could mean that
after these particles interact they may re-accelerate further, having their spectra pushed up in energy into the UHECR energy regime, 
retaining their spectral index,  before they start propagate.
The re-acceleration could happen in a nearby source like CenA, through relativistic shock acceleration,
where these particle spectra could be pushed up in energy before reaching Earths observatories, as Biermann and de Souza (2011) showed.

Alternatively,  assuming there could be a single source of very high energy particles 
within $d \simeq 50$ Mpc (note that Cen A is within 3-5 Mpc and M87 is within 16-17 Mpc), 
given our results, one could require additional losses in 
order to steepen the spectra accordingly,
and this could be feasible if the population of particles  
would be a mixture of protons and heavier nuclei (e.g. Fe), see Aloisio et al. (2011), Biermann and de Souza (2012).  
It is interesting to note here that the distance of either Cen A or M87 and the interaction of protons with the 
microwave background is not quite so important, see Greisen (1966), Zatsepin and Kuzmin (1966), except to the highest of energies.

On the other hand, if there were more that one source of very high energy cosmic rays (above $\sim 10^{9}$ GeV), within 50Mpc, 
(which we do not see yet) due to energy problems we described in Section 1, one would require a superposition of their spectra, 
as an integrated signal over the different shock profiles shown in this work, to satisfy the 
aforementioned calculations.
It is interesting to note here that of course there maybe many very high energy particle extragalactic sources,  
but it is hard to believe that they could all be sources of only heavy nuclei. 

One sees that our present study seems to agree with latest Auger data measurements, in connection to a proton composition in 
lower energies and probably a heavier composition imprint above $\sim 10^{10}$GeV.
It is important to note that our model results in spectra for which can reach a maximum of 
$\sim 10^{11}$ GeV for only one case, while the rest reach energies below $10^{10}$GeV, therefore justifying the first half 
of the UHECRs energy scale. For higher attained energies (up to $10^{12}$ GeV), higher Lorentz gamma shock factors than $\sim 50$ may be 
needed as it it shown in Meli et al. (2008), but for AGN jets plasma velocities this seems unlikely, except for some few cases e.g. 3C279.

The present acceleration model could  explain the variability of flat or inverted gamma-ray spectra observed in 
high red-shifted flaring extragalactic sources (e.g. 3C279; The MAGIC coll., 2008). These spectra may originate from particles accelerating and radiating 
from within a region of the topology we investigated, resulting in characteristic starved spectra shown in this work.

It seems possible that the population of UHECR is not only protons for the
whole energy spectrum examined here, but it can be a mixture of heavy nuclei and protons. 
If this is the case then it seems logic that a mixture of particles accelerating
in a shock sequence as the one described here, could give steeper spectra, given
heavy nuclei suffer photo-disintegration or fragmentation, for the highest energy spectral end.
This condition could change the spectral signatures considerably. This is a case we would like to examine further
in the near future.



\section{Conclusions}

We performed Monte Carlo simulations allowing acceleration at relativistic \textit{multiple} oblique
shocks as an application to muliple shocks observed in various AGN jets. We based our work on the re-confinement 
mechanism in jets and incomplete Comptonisation effect, injecting test-particles once upstream the first shock, and we 
let them accelerate through a sequence of subluminal and superluminal shocks. 

Specifically, we studied an exemplary case 
for six possible combinations of a set of four consecutive relativistic oblique shocks, 
with an aim to facilitate our understanding on acceleration efficiency and potential  particle spectra.
In principle, our investigation was initiated by the fact that particle energies with $\ge 10^{18.5}$ eV imply 
a lot more total source power. We showed that when flat and starved cosmic ray spectra are attainable, then the puzzling extra power to be 
provided for UHECRs does not seem necessary. 
By injecting particles once towards a sequence of shocks 
it means that with a smaller number of low energy particles one could have a final particle spectrum 
extending up to very high energies with depletion at lower energies.
We numerically showed that the first shock from the multiple-shock sequence, establishes 
a low energy power-law spectrum with $\sim E^{-2.7}$,  while the next consecutive shocks push the
particle spectrum up in energy with flatter distributions, leaving a flux deficiency at low energies.

We have shown that for two or more shocks the spectrum becomes flatten than -2.2. 
Especially, the case where only identical subluminal shocks are involved, the spectra are the flattest and the highest maximum particle
energy of a spectacular $10^{11}$eV 
is attained, confirming as well the expected high efficiency of subluminal
shocks shown previously by Meli \& Quenby (2003b), Meli et al. (2008). On the contrary when only superluminal shocks are present, 
they render the  spectra steeper and the acceleration less efficient. 

Within our model, other particles except protons (e.g. electrons, heavy nuclei), would naturally suffer 
losses in addition to all the effects from oblique shocks that could accentuate the sharpness of the final spectrum, 
as we discussed above, which is of interest to our astrophysical interpretation. 

Consequently, taking under consideration the work of de Marco \& Stanev (2005) and 
Berezinsky et al. (2006, 2009) which require an injected source spectrum for UHECRs between $E^{-2.4}$ and $E^{-2.7}$ before propagation,
our model can explain the latter on the basis of  either a single source such as Cen A, accelerating a composite population of protons and 
heavy nuclei, initially flat at source (justifying source power requirements) with additional propagation energy losses and a significant 
steepening, or 2),
due to the superposition of several sources, all of which end their acceleration shock sequence with two or 
more relativistic subluminal shocks with \textit{starved} flat spectra (justifying as well source power requirements).

Moreover, with the current model we could also explain  
irregular or very flat gamma-ray spectra by various flaring high red-shifted extragalactic sources. 
Specifically, inverted spectra may be
required to comprehend the tendency to detect TeV sources at redshifts
such as 0.5 (3C279; The MAGIC coll., 2008); sources with normal spectra would be
undetectable. This remains to be examined in more detail in a follow-up work.
Intensive future observations may yet to test or reveal one or more of the above predictions.



\begin{acknowledgements}
We acknowledge fruitful discussions with Prajval Shastri (Indian Institute of Astrophysics, Bangalore).
Support for work with PLB has come through the AUGER membership.
\end{acknowledgements}

\begin{appendix}

\section{Turbulence and scattering}

In standard kinetic theory the spatial diffusion coefficients $\kappa_{\|}$ and  $\kappa_{\perp}$, are related to the formula
$\kappa_{\perp}= \kappa_{\|}\cdot(1+(\lambda/r_g)^2)^{-1}$ (Ellison et al. 1995), where $\lambda$ is particle's m.f.p. 
The ratio $\lambda/r_g$ represents the level of media turbulence, where $r_{g}$ is the Larmor radius, in the plane of gyration.
In parallel shocks, where the magnetic field B field is directed
along the shock normal (i.e. $\psi=0^{o}$ ),  the ratio $\lambda/r_g$ has 
limited impact on the produced energy spectrum, which in principal determines the diffusive spatial scale normal to
the shock. Nevertheless, in oblique relativistic shocks ($\psi > 0^{o}$), 
the diffusive transport of particles across the shock, or in other words across the field,
becomes critical in retaining them into the acceleration process. Therefore, the interplay between the 
field angle and the value of the ratio $\lambda/r_g$ determines
the spectral index of the particle distribution (see also Baring 2004, Ellison and Double 2004).

Since a shock acceleration process implies that that energy gain rate scales as the gyrofrequency $c/r_{g}$, 
i.e. is proportional to the particle’s energy/momentum, in our work, a simple representation for the effect 
of the turbulence  is to suppose the particle scatters $\delta \theta=M/\Gamma$ every $\lambda= N r_{g}$
where $M$ and $N$  chosen numbers.
In the well known Bohm Limit, $\lambda/r_g =1$, where the field fluctuations satisfy  $\delta B/B \sim 1$. 
Interplanetary particle propagation studies and gyroresonance
theory suggest $\lambda$ to be a number of particle gyroradii $r_g$, that is $\lambda\geq 10 r_g$.
Moreover, in the shock normal, or in the $x$ direction, 
the diffusion coefficient can be written as $\kappa_{\|}=\lambda u /3$, when no cross-field diffusion occurs. 
When cross-field diffusion is present due to increased turbulence, then 
$\kappa=\kappa_{\|}\cos^{2} \psi + \kappa_{\perp}\sin^{2}\psi$ (Jokipii, 1987), where $\psi$ denotes the inclination
of the shock normal to the magnetic field.
For simplicity though it is generally assumed throughout the bibliography that $\kappa_{\|} >> \kappa_{\perp}$, so then a guiding center 
approximation can more accurately be applied.


Additionally, it is assumed that $\lambda$ is proportional to a power of the particle's momentum $p$ (Giacalone, et al. 1992) 
and it is presumed to scale as the particle gyroradius $r_g$ , i.e. $\lambda = N r_g \propto p$. 
When one reduces $\lambda/r_{g}$ then ones increases turbulence by allowing a very small degree 
cross-field diffusion. In its turn this may in general steepen the spectrum.  For values of $\lambda /r_g <10$ one may 
assume that qualitatively the shock behaves more like a parallel subluminal shock in terms of its diffusive character.

The so so-called large angle scattering is when particles scatter as $1/\Gamma \ll \delta\theta \leq \pi$, 
while in all other cases one has
a pitch angle diffusion, with extreme cases as $\delta\theta \leq 1/\Gamma$ (where $\Gamma$ denotes the Lorentz factor of the flow).
We mention here that in the work of Gallant et al. (1999) for relativistic shocks, it was demonstrated analytically that particles 
entering the upstream region in a direction nearly normal to the shock can only experience pitch angle diffusion of 
values $\delta \theta \leq 1/\Gamma$, with $\delta\theta$ measured 
in the upstream fluid frame for scattering in a uniform  field or a randomly oriented set of 
uniform field cells. This condition arises because particles attempting to penetrate upstream 
from the shock are swept back into the shock before they can scatter far from it. Nevertheless this is not entirely true, 
since the work of Gallant et al. (1999) was constrained only to the parallel shock case (see also Ellison et al., 1995, Baring, 2009). 
It is obvious that for oblique shocks, the kinematics involving particle
diffusion, turbulence ratio and the specific shock inclinations, play a critical role affecting 
the acceleration efficiency and particle spectra, as we will show below. This would be especially relevant, if the angle of 
the shock relative to the overall jet axis were to approach $1/\Gamma$.

It is true that in relativistic shocks the achieved energy constraint of particles is largely dependent on the 
kinematic competition between upstream particle flow and the relativistic approach of the shock front, on the other hand though
it is not critically dependent on the exact magnitude of the pitch angle scattering, as it was e.g. extensively discussed 
in Meli et al. (2008). One can allow pitch angle scattering greater than the extreme aforementioned case 
of $\delta \theta \leq 1/\Gamma$.
Furthermore, remembering it is the downstream scattering that is relevant
to particle loss, it seems reasonable to use larger values of $\delta\theta$  within the pitch angle scattering model.

A transverse field perturbation changes the pitch angle in a quasi-linear theory (Kennel 1966) by 

\begin{equation}
\delta \theta =\omega\frac{b}{B}\delta t\,
\end{equation}

in a time $ \delta t$ due to a perpendicular perturbation, $b$, to the mean field,  $B$, with cyclotron angular frequency,
$\omega= eB/\gamma m_{\circ}c$. The particle moves in near gyro-resonance with the wave in $b$.
A pitch angle diffusion coefficient can then be derived
\begin{equation}
D_{\theta}=\frac{\delta\theta^{2}}{\delta t}=\frac{\omega^{2}}{u_{\|}}\frac{P(k)}{B^{2}} \,,
\end{equation}

where $P(k)=P_{\circ}k^{s}$ is the power spectral density of $b$ at gyro-resonance wave number $k=\omega/u_{\|}$.
A particle then diffuses in pitch by a finite amount, $\delta \theta$ during $\delta t$  given by

\begin{equation}
\delta \theta^{2}=2D_{\theta}\delta t \,.
\end{equation}

It is waves, of wave number $k$, fulfilling the gyro-resonance condition, $k=\omega/u_{\|}$ that cause the scattering
of particles satisfying $\omega r_{g}=u_{\perp}$.  In the present simulations we also allow the particles with pitch angle chosen at
random to lie in the range of  $1/\Gamma \leq \delta \theta \leq 10/\Gamma$. On average a particle scatters 
$5/\Gamma$ after a time $10\sqrt{3}r_{g}/c$ so that

\begin{equation}
\delta \theta^{2}=\frac{25}{\Gamma^{2}}=\frac{2\omega^{2}}{u_{\|}B^{2}}P_{\circ}(k)\frac{\omega^{s}}  {u_{\|}^{s}}
\frac{10\sqrt{3}u_{\perp}}{c\,\omega}\,.
\end{equation}

To choose $\delta \theta$ independent of particle $\gamma$ that is of
$\omega$, we require $s$=-1 for the spectral slope. Then we obtain a power spectrum relative to the mean field power
\begin{equation}
\frac{P(k)}{B^{2}}=\frac{5}{4\sqrt{2}\Gamma^{2}}k^{-1}\,.
\end{equation}

The total fractional power in the turbulence if resonating waves are present to scatter particles of between 
$\gamma=300$ and $\gamma=10^{12}$ is $1.4/\Gamma^{2}$. Hence, the chosen pitch angle scattering model corresponds to
a mild turbulence situation, so accordingly we choose here $N=10$ that is, $\lambda= 10 r_{g}$.


Because quasi-linear theory for wave particle interactions is known to be an inexact
approximation, the power spectrum we have presented must also be an approximation to the scattering model we employ.
The choice of a fixed factor, $N=10$, in the relation between $\lambda$ and $r_{g}$ acknowledges this inexactness.  
The above discussion is presented to provide a link with the work of others, especially  Niemec \& Ostrowski (2005) 
who represent field fluctuations by employing predefined wave spectra in their studies. 

\end{appendix}

\end{document}